\documentclass[reprint,amsmath,amssymb,aps,pre,floatfix]{revtex4-2}

\usepackage{amsmath}
\usepackage{amssymb}

\usepackage{graphicx}               
\usepackage{wrapfig}

\usepackage{dcolumn}                

\usepackage{upgreek}                
\usepackage{bm}                     

\usepackage{xcolor}
\usepackage{hyperref}               

\usepackage{accents}
\usepackage{wasysym}                

\usepackage{lineno}
\usepackage{mathrsfs}
\usepackage{comment}
\usepackage{mathtools}
\usepackage[version=4]{mhchem}
\usepackage{float}
\usepackage{enumitem}
\usepackage{tikz}

\usepackage[normalem]{ulem} 

\hypersetup{colorlinks,linkcolor={blue!90!black},citecolor={blue!90!black},urlcolor={blue!90!black}}

\newcommand*\udot{{\mkern 1mu\cdot\mkern 1mu}}

\newcommand\um{{\textmu}m}  
\newcommand\uE{~{\textmu}E}  

\begin{document}


\title{Adaptive phototaxis of {\it Chlamydomonas} and the evolutionary\\ transition 
to multicellularity in Volvocine green algae}

\author{Kyriacos C Leptos}
\email[For correspondence:]{K.Leptos@damtp.cam.ac.uk}
\author{Maurizio Chioccioli}
\email[]{maurizio.chioccioli@yale.edu}
\altaffiliation[Present address: ]{Pulmonary, Critical Care and Sleep Medicine, Department of Internal Medicine, Yale School of Medicine, 300 Cedar Street, New Haven, CT 06519, USA}
\author{Silvano Furlan}
\email[]{silo.furlan@gmail.com}
\altaffiliation[Present address: ]{Sensing Electromagnetic Plus Corp., 2450 Embarcadero Way, Palo Alto, CA-94303, USA}
\author{Adriana I Pesci}
\email[]{A.I.Pesci@damtp.cam.ac.uk}
\author{Raymond E Goldstein}
\email[For correspondence:]{R.E.Goldstein@damtp.cam.ac.uk}
\affiliation{Department of Applied Mathematics and Theoretical Physics, University of Cambridge, Wilberforce Road, Cambridge, CB3 0WA, UK}

\date{\today}
\begin{abstract} 
A fundamental issue in biology is the nature of evolutionary transitions from unicellular to multicellular organisms.
Volvocine algae are models for this transition, as they span from
the unicellular biflagellate {\it Chlamydomonas} to multicellular species of {\it Volvox} with
up to 50,000 {\it Chlamydomonas}-like cells on the surface of a spherical
extracellular matrix.  The mechanism of phototaxis 
in these species is of particular interest since they lack a
nervous system and intercellular connections;  steering is
a consequence of the response of individual cells to light.  Studies of {\it Volvox}
and {\it Gonium}, a $16$-cell organism with a plate-like structure, have shown that the
flagellar response to changing illumination of the cellular photosensor is adaptive, with 
a recovery time 
tuned to the rotation period of the colony around its primary axis.  
Here, combining high-resolution studies of the flagellar photoresponse
with 3D tracking of freely-swimming cells, we show that such tuning also underlies 
phototaxis of {\it Chlamydomonas}.   A mathematical
model is developed based on the rotations around 
an axis perpendicular to the flagellar beat plane that occur through the 
adaptive response
to oscillating light levels as the organism spins. Exploiting a separation of time scales between
the flagellar photoresponse and phototurning, we develop
an equation of motion that accurately describes the observed photoalignment.
In showing that the adaptive time scale is tuned to the organisms' rotational period across three orders of magnitude in
cell number, our results suggest a
unified picture of phototaxis in green algae in which the asymmetry in torques
that produce phototurns arise from  
the individual flagella of {\it Chlamydomonas}, the flagellated edges of {\it Gonium} 
and the flagellated hemispheres of {\it Volvox}.
\end{abstract}

\maketitle

\section{Introduction}
\label{sec:intro}

A vast number of motile unicellular and multicellular eukaryotic microorganisms exhibits 
phototaxis, the ability to steer toward a light source, without possessing 
an image-forming optical system.  From photosynthetic algae \citep{Bendix1960} 
that harvest light energy to support their metabolic activities to 
larvae of marine zooplankton \citep{Thorson1964} whose upward phototactic motion
enhances their dispersal, the light sensor in such organisms is a single unit akin to one pixel
of a CCD sensor or one rod cell in a retina \cite{Hegemann_vision}.
In zooplanktonic larvae there is a single rhabdomeric photoreceptor cell \citep{Jekely2008} while 
motile photosynthetic microorganisms such as green algae \cite{Hegemann_review}
have a ``light antenna" \cite{Foster1980}, which co-localizes with a cellular structure 
called the \textit{eyespot}, a carotenoid-rich orange stigma. 
For these simple organisms, the process of {\it vectorial phototaxis}, 
motion in the direction of a source rather than in response to 
a light gradient \cite{Giometto},  relies on an interplay between the detection of light 
by the photosensor and changes to the actuation of the apparatus that confers motility, namely their one or 
more flagella. Evolved independently many times \cite{Jekely_evolution}, the common sensing/steering mechanism 
seen across species involves two key features.  

The first attribute is a photosensor  
that has {\it directional} sensitivity, detecting only light incident from one side.  
It was hypothesized long ago \cite{Foster1980} that in green algae this
asymmetry could arise if the layers of carotenoid vesicles behind the
actual photosensor act as an interference reflector. In zooplankton this ``shading" role 
is filled by a single pigment cell \citep{Jekely2008}.
This directionality hypothesis was verified in algae by experiments on 
mutants without the eyespot, that lacked the carotenoid vesicles \citep{Ueki2016}, so that light 
could be detected whatever its 
direction. Whereas wild-type cells performed positive phototaxis (moving toward
a light source), the mutants might naively have been expected to be incapable of phototaxis.
Yet, they exhibited {\it negative} phototaxis, a fact that was explained as a consequence
of an additional effect first proposed earlier \cite{Kessler_lens}; the algal cell body functions as a
convex lens with refractive index greater than that of water.  Thus, a
greater intensity of light falls on the photosensor when it was illuminated from behind
than from the front, and a cell facing away from the light erroneously continues swimming in 
that direction, as if it were swimming toward the light.

The second common feature of phototactic microorganisms is a natural swimming trajectory
that is helical.
Spiral swimming has been remarked upon since at least the early 1900s, when Jennings \cite{Jennings} suggested
that it served as a way of producing trajectories that are straight on the large scale, while compensating for
inevitable asymmetries in the body shape or actuation of cilia, and Wildman \cite{Wildman} presciently observed
that chirality of swimming and ciliary beating must ultimately be understood in terms of the genetic program
contained within chromosomes.  While neither offered a functional purpose related to phototaxis,
Jennings did note earlier \cite{Jennings_early1,Jennings_early2} that when organisms swim along regular helices 
they always present the same side of their body to the outside.  This implies that during
regular motion the photosensor itself also has a fixed relationship to the helix.  
 
\begin{figure}[t]
\includegraphics[width=0.98\columnwidth]{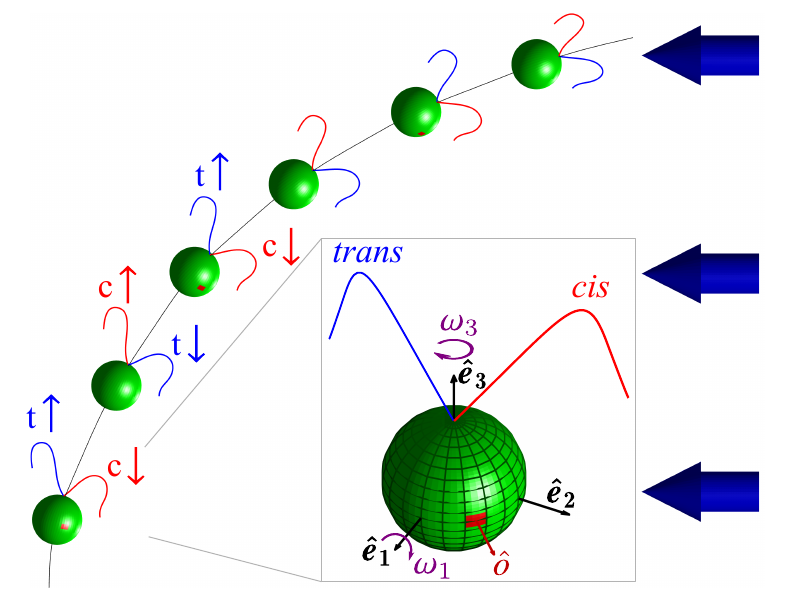}
\caption{Phototaxis in \textit{Chlamydomonas}.  Cell geometry (in box) involves the primary
axis $\hat{\bf e}_3$ around which the cell spins at angular frequency $\omega_3$, the
axis $\hat{\bf e}_2$ in the flagellar beat plane, and 
$\hat{\bf e}_1=\hat{\bf e}_2\times \hat{\bf e}_3$.
As the cell swims and spins around $\hat{\bf e}_3$, its
eyespot (red) moves in an out of the light shining in the direction of the blue
arrows.  This periodic light stimulation 
leads to alternating {\it cis} and
{\it trans} flagella dominance, producing rotations around $\bf{\hat{e}_1}$ and hence a
phototactic turn.}
\label{fig1}
\end{figure}

In {\it Chlamydomonas},
motility derives from the breaststroke beating of two oppositely-oriented flagella emanating from
near the anterior pole of the cell body, as depicted in 
Fig.~\ref{fig1}.  The flagella, termed {\it cis} and {\it trans} for their 
proximity to the eyespot, define a plane, the unit normal to which is the vector $\hat{\bf e}_1$. 
Historical uncertainties 
around the precise three-dimensional swimming motion of {\it Chlamydomonas} 
were resolved with the work of Kamiya and Witman \cite{KamiyaWitman}, the high-speed imaging study of 
R{\"u}ffer and Nultsch \cite{Ruffer1985} and later work by Schaller \textit{et al.} \cite{Schaller1997}, who
together demonstrated three features: (i) the eyespot is typically located on the equatorial plane of
the cell, midway between $\hat{\bf e}_1$ and the vector $\hat{\bf e}_2$ that lies within the 
flagellar plane, pointing toward the {\it cis} flagellum, (ii) cells rotate counterclockwise (when
viewed from behind) the axis
$\hat{\bf e}_3$ at frequency $f_{\mathrm{r}}\sim 1.5-2.5\,$Hz 
($f_r=1.67\pm 0.35\,$Hz in a recent direct measurement \cite{Choudhary}), and (iii) positively phototactic 
cells swim along helices such that the 
eyespot always faces 
{\it outward}.  The rotation around $\hat{\bf e}_3$ was conjectured to arise from a small
non-planarity of the beat, as has been recently verified \cite{Kirsty}, while
helical swimming arises from rotation around $\hat{\bf e}_1$ due to a slight asymmetry 
in the two flagellar beats. 

It follows from the above that the eyespot of a cell whose swimming is
not aligned to the light receives an oscillating signal 
at angular frequency $\omega_3 = 2\pi f_{\mathrm{r}}$. Detailed investigation into the effect of this periodic signal 
began with the work of 
R{\"u}ffer and Nultsch, who used cells immobilized on micropipettes to enable high-speed cinematography of the
waveforms.  Their studies \cite{Ruffer1990,Ruffer1991} of beating dynamics in 
a negatively-phototactic strain showed the
key result that the {\it cis} and {\it trans} flagella responded differently to changing light levels
by altering their waveforms in response to the periodic steps-up and steps-down in signals that occur 
as the cell rotates. 
This result led to a model for phototaxis \citep{Schaller1997} that divides turning into two phases 
(Fig.~\ref{fig1}): \textit{phase I}, in which the eyespot moves from shade to light, causing the 
{\it trans} flagellum to increase transiently its amplitude relative to the {\it cis} flagellum, and 
\textit{phase II}, in which the eyespot moves from light to 
shade, leading to transient beating with the opposite asymmetry.  Both phases lead to 
rotations around $\hat{\bf e}_1$, and turns toward the light.
The need for an asymmetric flagellar response was shown in studies of
the mutant {\it ptx1} \cite{RufferNultsch1997,Okita2005}, which lacks 
calcium dependent flagellar dominance \cite{Bessen} and 
can not do phototaxis.

These transient responses were studied further \cite{Yoshimura2001} 
through the 
photoreceptor current (PRC) that can be measured in the surrounding fluid.  Subjecting a 
suspension of immotile cells 
(chosen to avoid movements) to rectified sinusoidal light signals that mimic those received 
by a rotating cell,
they found that the PRC amplitude displays a maximum as a function of frequency, with 
a peak close to the body rotation frequency $f_r$.
This ``tuning" of the response curve was investigated in more detail\textemdash  in 
a negatively-phototactic strain\textemdash in 
the important work of Josef, et al. \cite{Josef2005},
who projected the image of the cell onto a quadrant photodiode whose analog signal could be digitized at up to
$4000$ samples per second.  While this device did not allow detailed imaging of the entire waveform, 
it was able to capture changes in the forward reach of the two flagella (termed the  ``front amplitude") 
over significantly longer time series than previous methods.
Combined with later work that analyzed the signals within the framework of linear systems 
analysis \cite{Josef2006}, these studies showed how each of the two flagella exhibits a distinct, 
peaked frequency response.

From the original measurements of transient PRCs induced by step changes in light
levels \cite{Yoshimura2001}, it was evident that
the response in time was biphasic and {\it adaptive} \textemdash a rapid
rise in signal accompanied by slower recovery phase back to the resting state \textemdash and the 
presence of two timescales is implicit in the existence of the peak in the 
frequency response.  More recently, measurements of the flagella-driven 
fluid flow around colonies of the
multicellular alga {\it Volvox carteri} \cite{Drescher2010} showed again this adaptive response, 
which could be described quantitatively by a 
model previously to describe chemotaxis of both 
bacteria \cite{Othmer} and spermatozoa \cite{Friedrich2007}.
In a suitably rescaled set of units, the two variables
$p$ and $h$ in this model respond to a signal $s(t)$ through the coupled ODEs%
\begin{subequations}
\begin{align}
\tau_r\dot{p}&=s-h - p \\
\tau_a\dot{h}&=s-h,    
\end{align}
\label{eq:model}%
\end{subequations}
where $p$ governs some observable, $h$ represents hidden biochemistry
responsible for adaptation, $\tau_r$ is the rapid response time, and $\tau_a$ is the
slower adaption time.  In bacteria, the adaptive response is
exhibited by the biochemical network governing rotation of flagella,
while for sperm curvature of the swimming path was altered
linearly with $p$ in response to a chemoattractant.

The model \eqref{eq:model} was incorporated into a theory of {\it Volvox} phototaxis 
using a coarse-grained description of flagella-driven
flows akin to the squirmer model \cite{Lighthill}, with 
a dynamic slip velocity ${\bf u}(\theta,\phi,t)$ as a function of 
spherical coordinates on the 
colony surface.  Without light stimulation, the velocity is an axisymmetric function 
${\bf u}_0(\theta)$ that varies with the polar angle $\theta$, and is dominated by the 
first mode $u_1\propto \sin\theta$ \cite{Short2006}.
As \eqref{eq:model} is meant to describe the fluid flow associated
with flagella of each of the somatic cells on the surface, it is introduced
into the slip-velocity model through response {\it fields} $p(\theta,\phi,t)$ and $h(\theta,\phi,t)$ 
over the entire surface.
Experiments indicate
that the photoresponse is accurately represented the form
\begin{equation}
    {\bf u}(\theta,\phi,t)={\bf u}_0(\theta)\left[1-\beta(\theta)p(\theta,\phi,t) \right],
\end{equation}
where the parameter $\beta$ encodes the latitude-dependent photoresponse of the flagella 
(strong at the
anterior of the colony, weak in its posterior).
The swimming trajectories were then obtained from integral relationships between the slip velocity
and the colony angular velocity \cite{StoneSamuel}.  

\begin{figure}[t]
\includegraphics[width=0.98\columnwidth]{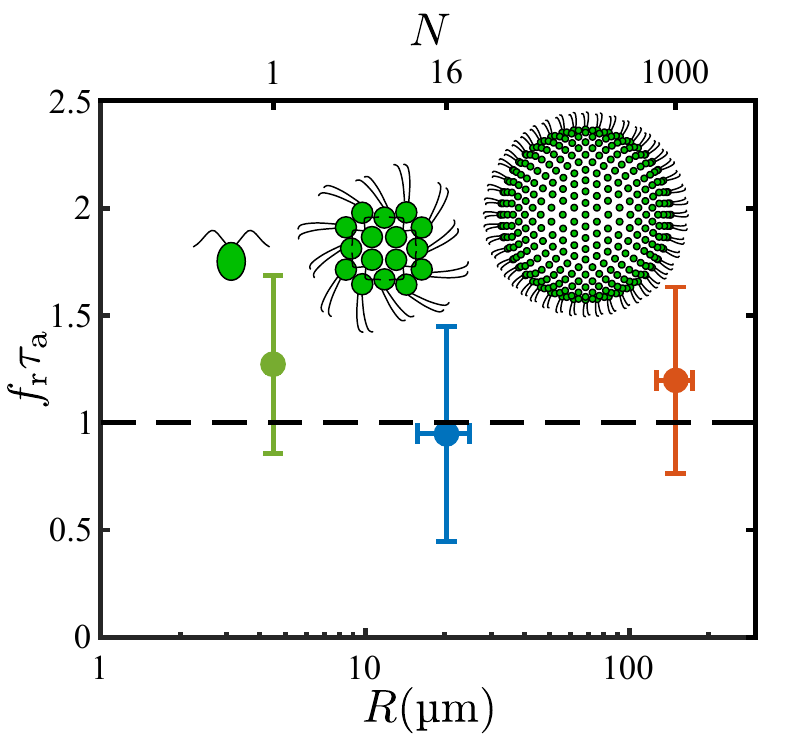}
\caption{Master plot of adaptive time scales in Volvocine green algae.  For each of 
{\it Chlamydomonas} (this paper), {\it Gonium} \cite{GoniumPRE}, and {\it Volvox} 
\cite{Drescher2010} the dimensionless product of the rotation frequency $f_r$ around the primary body-fixed
axis and the flagellar 
adaptive time $\tau_a$ is plotted as a function of the
organism radius $R$ (bottom axis) and typical cell number $N$ (top axis).}
\label{fig2}
\end{figure}

Statistical analysis of many {\it Volvox} colonies shows that
there is tuning of the response in that the product $f_r\tau_a\approx 1$ 
($f_r\tau_a=1.20\pm 0.44$) \cite{Drescher2010}, as indicated
in Fig.~\ref{fig2}. The significance of the product $f_r\tau_a$ being of order unity can be
understood as follows:  when a region of somatic cells rotates 
to face a light source, the fluid flow it produces will decrease as $p$ rapidly increases on a time
scale $\tau_r$, and if the time $\tau_a$ it takes for $p$ to recover is comparable 
to the colony rotation period $1/f_r$ then the fluid flow along the dark side 
will be stronger than than on the light side, and the colony turns to the light. 

A similar tuning phenomenon is found with {\it Gonium} \cite{GoniumPRE}, a member of the Volvocales 
typically composed of $16$ cells arranged in a flat
sheet as in Fig.~\ref{fig2}.  The flagella of the four
central cells beat in a {\it Chlamydomonas}-like breaststroke waveform that 
propels the colony in the direction of the body-fixed axis $\hat{\bf e}_3$ 
perpendicular to the sheet.  The flagella of the outer $12$ cells beat at an angle
with respect to the plane; their dominant in-plane component rotates 
the colony at frequency $f_r$ about $\hat{\bf e}_3$, while the out-of-plane component
adds to the propulsive force of the central cells. Experiments show that the
peripheral cells display the same kind of 
biphasic, adaptive response as do {\it Volvox} colonies.  This light-induced 
``drop-and-recover response" produces 
an axial force component $f_\parallel$ from the peripheral flagella 
of the form
\begin{equation}
    f_\parallel(\theta,t)=f_\parallel^{(0)}\left[1-p(\theta,t) \right],
\end{equation}
where $f_\parallel^{(0)}$ is the uniform component in the absence of photostimulation. 
Again, the directionality of the eyespot sensitivity leads to a photoresponse $p$ that is
greatest (and $f_\parallel$ that is smallest) for those cells facing the light, and this 
nonuniformity in $f_\parallel$ leads to 
a net torque about an in-plane axis which, balanced by rotational drag, leads to phototactic
turning toward the light.  The data for {\it Gonium} also supports
tuning, with the product $f_r\tau_a=0.95\pm 0.50$, as shown in Fig.~\ref{fig2}.
 
In the present work we complete a triptych of studies in Volvocine algae 
by examining {\it Chlamydomonas}, the unicellular ancestor of all others \cite{biorxiv}.
Our purpose is to construct, in a manner that parallels that for {\it Volvox} and {\it Gonium}, a 
theory 
that links the photoresponse
of flagella to the trajectories of cells turning to the light. 
We base the description on the kinematics of rigid bodies, where the central quantities are the angular velocities 
around body-fixed axes. 
This model bears some similarity to an earlier study of phototaxis \citep{Bennett2015}, in which the asymmetric 
beating of flagella\textemdash modelled as spheres moving along orbits under the action of prescribed
internal forces responding to light on the eyespot\textemdash was related to rotations about
body-fixed axes, but the response to light was taken to be instantaneous and non-adaptive.

Results reported here on {\it Chlamydomonas} show that $f_r\tau_a$ is
close to unity ($f_r\tau_a=1.27\pm 0.41$), from which we infer that tuning is an 
evolutionarily conserved feature spanning three orders of magnitude in cell number and nearly 
two orders of magnitude in organism radius (Fig.~\ref{fig2}).  We conclude that, in evolutionary 
transitions to multicellularity
in the Volvocine algae, the ancestral photoresponse found in 
{\it Chlamydomonas} required little modification in order to work in vastly 
larger multicellular spheroids.  The most
significant change is basal body rotation \cite{Kirk2005} in the multicellulars in order
that the two flagella on each somatic cell beat in parallel, rather than opposed as in {\it Chlamydomonas}.
In {\it Gonium}, this arrangement in the peripheral cells leads to colony rotation, while for
the somatic cells of {\it Volvox} the flagellar beat plane is tilted with respect to meridional lines,
yielding rotation around the primary colony axis.

The presentation below proceeds from small scales to large, following 
a description in Sec. \ref{sec:methods} of experimental methods used in our 
studies of the flagellar photoresponse of 
immobilized cells at high spatio-temporal resolution, and of methods for
tracking phototactic cells.  In Sec. \ref{sec:flag_results} we 
arrive at an estimate of rotations about the body-fixed axis $\hat{\bf e}_1$
arising from transient flagellar asymmetries induced by light falling on the eyespot,
and thus a protocol to convert
measured flagella dynamics to angular velocities within the adaptive model.
Section \ref{sec:turndynamics} incorporates those results into a theory of phototactic turning.
Exploiting a separation of time scales between individual flagella beats, cell rotation, and 
phototactic turning, we show how the continuous-time dynamics can be approximated
by an iterated map, and allow direct comparison to three-dimensional trajectories 
of phototactic cells.  By incorporating an adaptive dynamics
at the microscale, one can examine the speed and stability of phototaxis as a function
of the tuning parameter $f_r\tau_a$ and deduce its optimum value.
These results explain the many experimental
results summarized above, and put on a mathematical basis phenomenological arguments
\cite{Schaller1997} about the stability of phototactic alignment in 
{\it Chlamydomonas}.  

\begin{figure}[t]
\includegraphics[width=0.98\columnwidth]{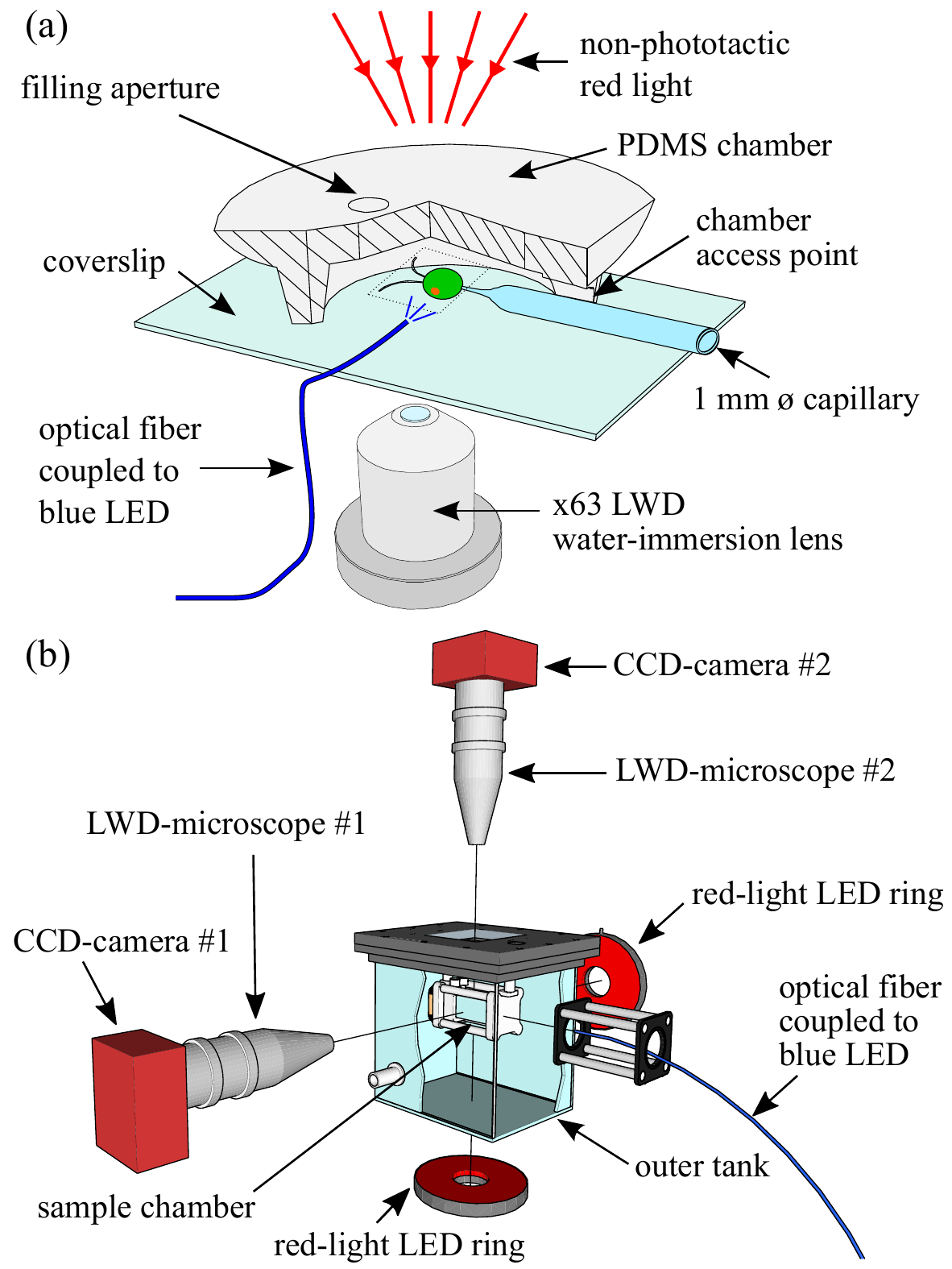}
\caption{Experimental methods.
Setups to measure (a) the flagellar photoresponse of cells immobilized on a micropipette, 
and (b) swimming trajectories of phototactic cells in a sample chamber 
immersed in an outer water tank to minimize thermal convection, as described in text.}
\label{fig3}
\end{figure}

\section{Experimental methods}
\label{sec:methods}

{\it Culture conditions}.  Wild type
\textit{Chlamydomonas reinhardtii} cells (strain CC125 \citep{chlamy_sourcebook}) were grown axenically 
under photoautotrophic conditions in minimal media \citep{rochaix}, at 23$^{\circ}$C under 
a 100\uE$\cdot$s$^{-1}\cdot$m$^{-2}$ illumination in 
a diurnal growth chamber with a $14:10\,$h light-dark cycle.

{\it Flagellar photoresponse of immobilized cells.}
The flagellar photoresponse of \textit{C. reinhardtii} was captured at high spatio-temporal resolution 
using the experimental setup shown in Fig.~\ref{fig3}(a), which builds on 
previous studies \citep{Polin2009, Drescher2010, Leptos2013}.
Cells were prepared as described previously \citep{Leptos2013} -- centrifuged, washed and gently 
pipetted into a bespoke observation chamber made of polydimethylsiloxane (PDMS). Chambers were mounted on a 
Nikon TE2000-U inverted microscope with a $\times63$ Plan-Apochromat water-immersion long-working-distance (LWD) 
objective lens (441470-9900; Carl Zeiss AG, Germany). Cells were immobilized via aspiration using a 
micropipette (B100-75-15; Sutter, USA) 
that was pulled to a \diameter 5-\textmu m tip, and the flagellar beat plane was aligned with the focal plane of the 
objective lens via a rotation stage. Video microscopy of immobilized cells was performed using a high speed camera 
(Phantom v341; Vision Research, USA) by acquiring $15\,$s movies at $2,000$ fps. 

The light used for photostimulation of cells was provided by a 
$470$ nm Light Emitting Diode (LED) (M470L3; Thorlabs, USA) that 
was controlled via an LED driver 
(LEDD1B; Thorlabs, USA), coupled to 
a \diameter $50\,$\um-core optical fiber (FG050LGA; Thorlabs, USA).  This fiber is much smaller than 
that used in previous versions of this setup in order to accommodate 
the smaller size of a \textit{Chlamydomonas} cell relative to a \textit{Volvox} spheroid.
The LED driver and the high-speed camera were triggered through a data-acquisition card 
(NI PCIe-6343; National Instruments, USA) using in-house programs written in \texttt{LabVIEW 2013} 
(National Instruments, USA), for both step- and frequency-response experiments. 
Calibration of the optical fiber was performed as follows: A photodiode (DET110; Thorlabs, USA) was used to 
measure the total radiant power $W$ emerging from the end of the optical fiber 
for a range of voltage output values (0-5 V) of the LED driver. The two quantities were plotted and 
fitted to a power-law model which was close to linear.

Cells were stimulated at frame 
$2896$ ($\approx 1.45\,$s into the recording).  A light intensity 
of $\approx$ 1\uE$\cdot$s$^{-1}\cdot$m$^{-2}$ 
(at $470\,$nm) was found empirically to give the 
best results in terms of reproducibility, sign, i.e. positive phototaxis, and quality of response; we conjecture 
that the cells could recover in time for the next round of stimulation. For the step-response experiments, 
biological replicates were $n_\mathrm{cells} = 3$ with corresponding technical replicates 
$n_\mathrm{tech} = \{4, 2, 2\}$. For the 
frequency-response experiments, biological replicates were $n_\mathrm{cells} = 3$ with each cell 
stimulated to the following amplitude-varying frequencies: 0.5 Hz, 1 Hz, 2 Hz, 4 Hz and 8 Hz. Only 
the cells that showed a positive sign of response for \textit{all} 5 frequencies are presented here. 
This was hence the most challenging aspect of the experimental procedure.

To summarize, the total number of high-speed movies acquired was $n_{\mathrm{mov[hs]}} = 24$. All downstream analysis of the movies was carried out in \texttt{MATLAB}. Image processing and flagella tracking was based on previous work \cite{Leptos2013}, and new code was written for force/torque calculations and flagellar photoresponse analysis.

{\it Phototaxis experiments on free-swimming cells.}
Three-dimensional tracking of phototactic cells was performed using the method described previously \cite{Drescher2009} 
with the modified apparatus shown in Fig.~\ref{fig3}(b).
The experimental setup comprised of a sample chamber suspended in an outer water tank to eliminate thermal convection.
The modified sample chamber was composed of two acrylic flanges (machined in-house) that were clamped in a watertight
manner onto an open-ended square borosilicate glass tube ($2\,\text{cm} \times 2\,\text{cm} 
\times 2.5\,\text{cm}$; Vetrospec Ltd, UK).  This design allowed a more accurate and easy calibration of the field 
of view and a simpler and better loading system of the sample via two barbed fittings. This 
new design also minimized sample contamination during experiments.
Two $6$ megapixel charge-coupled device (CCD) cameras (Prosilica GT2750; Allied Vision Technologies, Germany), 
coupled to 
two InfiniProbe\textsuperscript{TM} TS-160s (Infinity, USA) with Micro HM objectives were used
to achieve a larger working distance than in earlier work ($48\,$mm vs. $38\,$mm) 
at a higher total magnification of $\times16$.
The source of phototactic stimulus was a $470\,$nm blue-light LED (M470F1; Thorlabs, USA) coupled to 
a solarization-resistant optical fiber (M22L01; Thorlabs, USA) attached to an in-house assembled fiber 
collimator that included a \diameter 12.7 mm plano-convex lens (LA1074-A; Thorlabs, USA).
Calibration of the collimated optical fiber was performed similarly to the experiments with 
immobilized cells. The calibration took account of the thickness of the walls of the outer water tank
and the inner sample chamber, as well as the water 
in between.

The two CCD cameras and the blue-light LED used for the stimulus light were controlled using 
\texttt{LabVIEW 2013} (National Instruments, USA) including the image acquisition driver 
\texttt{NI-IMAQ} (National Instruments, USA).
The cameras were triggered and synchronized at a frame rate of $10\,$Hz via a data-acquisition device 
(NI USB 6212-BNC; National Instruments, USA).
For every tracking experiment ($n_{\mathrm{mov[3d]}} = 6$), two $300$-frame movies were acquired (side and top) with the 
phototactic light triggered at frame $50$ ($5\,$s into the recording).
The intensity of the blue-light stimulus was chosen to be either $5$ or 10\uE$\cdot$s$^{-1}\cdot$m$^{-2}$.
To track the cells we used in-house tracking computer programs written in \texttt{MATLAB} as 
described in \cite{Drescher2009}. Briefly, for every pair of movies cells were tracked in the 
\textit{side} and \textit{top} movies corresponding to the $xz$-plane and in the $xy$-plane respectively. 
The two tracks were aligned based on their $x$-component to reconstruct the three-dimensional trajectories. The angle between the cell's directional vector and the light was then calculated for every time point.

\section{Flagellar dynamics}
\label{sec:flag_results}

\begin{figure}[t]
\includegraphics[width=0.98\columnwidth]{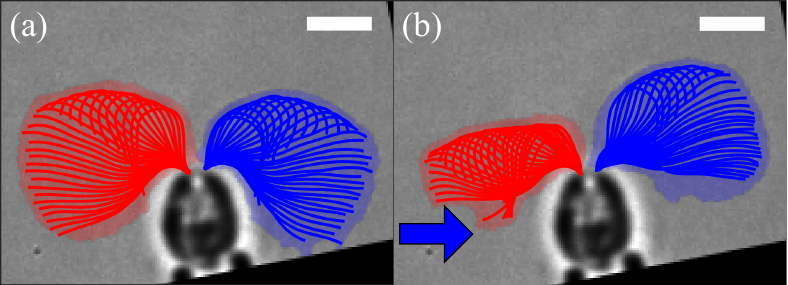}
\caption{Flagellar photoresponse of an immobilized cell after a step-up in light.
Light is from the left (blue arrow), toward the eyespot.  The panels show 
overlaid flagellar waveforms of
a single beat (a) in the dark and (b) starting at $50\,$ms ($52.5\,$ms) for the
{\it cis} ({\it trans}) flagellum after the step increase in light. 
Scale bar is $5\,$\um.
}
\label{fig4}
\end{figure}

\subsection{Forces and torques}
\label{cellbodymotion}

We begin by examining the response of the two flagella of an immobilized 
{\it Chlamydomonas} cell to a change in the light level illuminating the eyespot.  
Figure \ref{fig4} and Supplementary Video 1 \cite{SM} 
shows a comparison between the unstimulated beating of the flagella and the 
response to a simple step up from zero illumination.  These are presented
as overlaid flagellar waveforms during a the single beat in the dark and one that started $50\,$ms after
the step.
In agreement with previous work cited in Sec. \ref{sec:intro} \citep{Ruffer1991,Josef2005}, 
we see that the
transient response involves the \textit{trans} flagellum reaching further
forward toward the anterior of the cell, while the \textit{cis} waveform 
contracts dramatically.
The photoresponse is adaptive;
the marked asymmetry between the {\it cis} and {\it trans} waveforms decays away over $\sim 1-2\,$s,
restoring the approximate symmetry between the two.  This adaptive timescale is much 
longer than the period $T_\mathrm{b}\sim 20\,$ms of
individual flagellar beats.  

\begin{table}[t]
\caption{Geometry of flagellar beats.  Data are from the
present study except for the flagellum radius.}
\begin{ruledtabular}
\begin{tabular}{cccc}
\textbf{Quantity} & \textbf{Symbol} & \textbf{Mean} ${\bm{\pm}}$ \textbf{SD}\\
\hline 
flagellum length & $L$ & $13.5\,\pm\,0.8\,$\textmu m\\
flagellum radius \cite{Sager}& $a$ & $0.125\,$\textmu m\\
cell body radius & $R$ & $4.4\,\pm\,0.3\,$\textmu m\\
beat frequency & $\hat{f}_\mathrm{b}$ & $43.2\,\pm\,8.0\,$Hz\\
anchor angle & $\varphi_a$ & $(0.05\pm0.01)\pi$ \\
initial angle & $\hat{\varphi}_0$ & $(0.26\,\pm\,0.05)\pi$ \\
sweep angle & $\hat{\varphi}_\mathrm{b}$ & $(0.33\,\pm\,0.06)\pi$ \\
\end{tabular}
\end{ruledtabular}
\label{table1}
\end{table}

\begin{figure}[t]
\includegraphics[width=0.98\columnwidth]{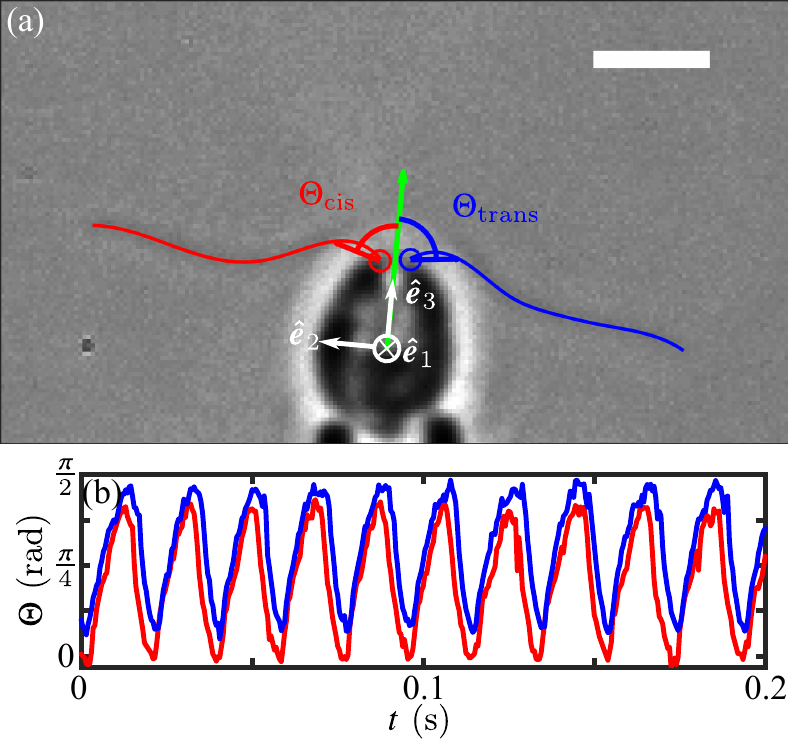}
\caption{Flagellar beat cycles.  (a) Angles $\Theta$ on each flagellum 
(red for \textit{cis}, blue for \textit{trans}) relative to symmetry axis 
$\hat{\bf e}_3$ (green) of the 
cell are used to define the cycle. Scale bar is $5\,$\um. (b) Typical time series 
of the two angles.}
\label{fig5}
\end{figure}

We wish to relate transient
flagellar asymmetries observed with immobilized cells, subject to time-dependent 
light stimulation, to cell rotations 
that would occur for freely-swimming cells.  We begin by examining the beating of unstimulated cells to provide
benchmark observations.  We analyze high-speed videos to obtain the waveforms of flagella of length $L$,
radius $a$,
in the form of moving curves 
${\bf r}(\lambda,t)$ parameterized by arclength $\lambda\in [0,L]$ and time.
Within Resistive Force Theory (RFT) \cite{RFT,Laugabook}, and specializing to planar curves, 
the hydrodynamic force density on the filament is
\begin{equation}
    {\bf f}(\lambda,t)=-\left(\zeta_\perp \hat{\bf n}\hat{\bf n}
    +\zeta_\parallel \hat{\bf t}\hat{\bf t}\right)\cdot {\bf r}_t(\lambda,t),
    \label{RFT}
\end{equation}
where $\hat{\bf t}={\bf r}_s$ and $\hat{\bf n}$ (with $\hat{\bf n}_i=\epsilon_{ji}\hat{\bf t}_j$) 
are the unit normal and tangent at $\lambda$, 
and $\zeta_\perp$ and $\zeta_\parallel$ are
drag coefficients for motion perpendicular and parallel to the filament.  
We assume the asymptotic results
$\zeta_{\perp}=4\pi\mu/c_{\perp}$ and $\zeta_\parallel=2\pi\mu/c_\parallel$, 
where $c_\perp=\ln({\cal L}\sqrt{e})$ and $c_\parallel=\ln({\cal L}/\sqrt{e})$, 
with ${\cal L}=L/a$ the aspect ratio.  Table \ref{table1} gives typical values of the 
cell parameters; with ${\cal L}\approx 108$, we have $c_\perp\approx 5.2$ and $c_\parallel\approx 4.2$.
To complete the analysis, we adopt the convention shown in
Fig.~\ref{fig5}(a) to define the start of a beat, 
in which chords drawn from the base to a point of fixed length on each flagellum 
define angles $\Theta_\mathrm{cis,trans}$ with respect to
$\hat{\bf e}_3$,  Local minima in $\Theta_\mathrm{cis,trans}(t)$
[Fig.~\ref{fig5}(b)] define the beat endpoints.

\begin{figure}[t]
\includegraphics[width=0.98\columnwidth]{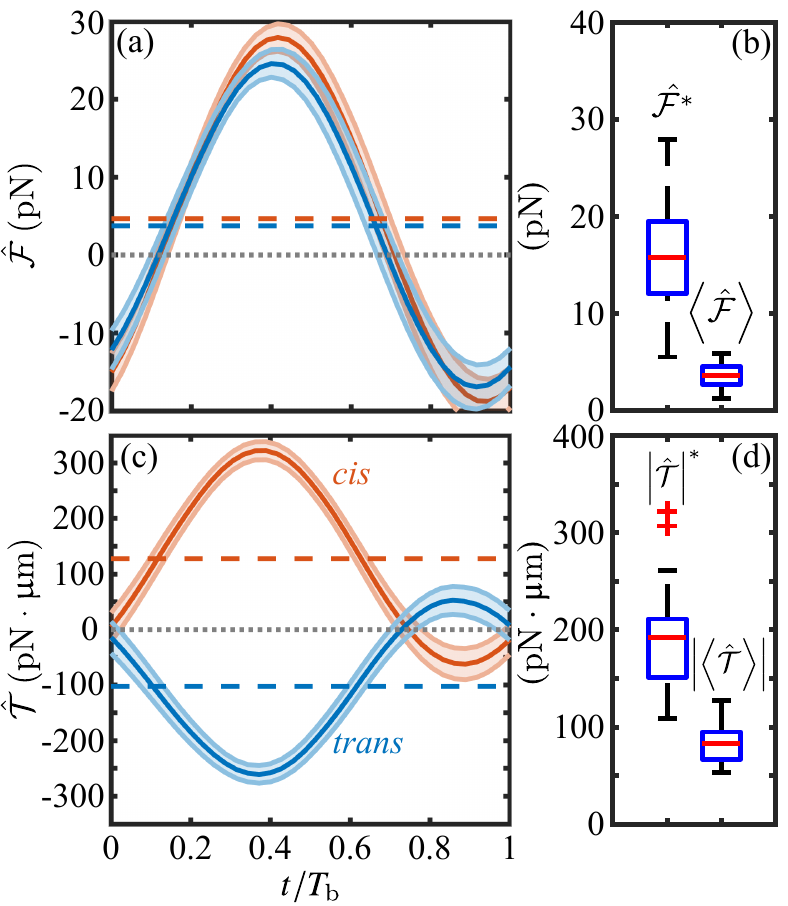}
\caption{Flagellar forces and torques of unstimulated cells. Propulsive force (a) and
torque about the cell center (c) of 
\textit{cis} (red) and \textit{trans} (blue) flagella during beat cycle of a
representative cell, with cycle
averages indicated by dashed lines.
(b) and (d) show boxplots of peak values ($^*$) and beat-average quantities ($\langle\rangle$) 
computed from $n = 48$ flagella in 24 movies.}
\label{fig6}
\end{figure}

Using a hat ($\,\hat{}\,$) to denote quantities 
measured without photostimulation, Fig.~\ref{fig6} shows the results 
of this analysis for the 
propulsive component of the total force,
\begin{equation}
    \hat{\cal F}(t)=\hat{\bf e}_3\cdot\int_0^L\! d\lambda\, {\bf f}(\lambda,t)
    \label{Fpropulsive}
\end{equation}
and the torque component around $\hat{\bf e}_1$,
\begin{equation}
    \hat{\mathcal T}(t)=\hat{\bf e}_1 \cdot\int_0^L\! d\lambda\, {\bf r}\times {\bf f}(\lambda,t),
\end{equation}
where ${\bf r}$ is measured from the cell center.  The smoothness of the data arises from
the large number of beats over which the data are averaged.
The force $\hat{\mathcal F}$ varies sinusoidally in time, offset from zero 
due to the dominance of the power stroke over the recovery stroke, with a peak 
value $\hat{\mathcal F}^*= 16.0\pm 5.2\,$pN and mean over a beat period of 
$\langle \hat{\mathcal F}\rangle=3.6\pm 1.1\,$pN per flagellum.
These findings are in general agreement with previous studies of {\it Chlamydomonas} cells held by an optical trap \cite{McCord}, micropipette-held 
cells \cite{Brumley2014}, measurements on swimming 
cells confined in thin fluid films \cite{Guasto2010} and in 
bulk \cite{Klindt2015}, and more 
recent work using using micropipette force sensors \cite{Boddeker}.

Aggregating all data obtained on the unstimulated torques exhibited by {\it cis} and 
{\it trans} flagella, we find a peak magnitude 
$\hat{\mathcal{T}}^*=187\pm 48\,$pN$\cdot$\textmu m
and cycle-average mean value $\langle \hat{\mathcal T}\rangle=82\pm 17\,$pN$\cdot$\textmu m.
As a consistency check we note that the ratio torque/force 
should be an interpretable length, 
and we find $\hat{\mathcal{T}}^*/\hat{\mathcal F}^*\sim 12\,$\textmu m, a value that is
very close to the mean flagellar length $L=13.5\,$\textmu m.  
Across a sample size of $n=24$ we find the sum 
\begin{equation}
    \langle \hat{\mathcal T}\rangle=\langle \hat{\mathcal T}_{\rm cis}\rangle 
    +\langle\hat{\mathcal T}_{\rm trans}\rangle,
    \label{residual_torque}
\end{equation}
is $\langle \hat{\mathcal T}\rangle=-6.2 \pm 15.9\,$ pN$\cdot$\textmu m, and thus
is consistent with symmetry of the two flagella, but the data 
clusters into two clear groups;
a {\it cis}-dominant subpopulation ($n=9$), with 
$\langle \hat{\mathcal T}\rangle^{\rm cis}=11.2 \pm 7.0\,$pN$\cdot\upmu$m and a
{\it trans}-dominant subpopulation ($n=15$) with
$\langle \hat{\mathcal T}\rangle^{\rm trans}=-16.6 \pm 8.6\,$pN$\cdot$\textmu m.
As discussed below, such differences would generally distinguish between positively and
negatively phototactic cells, and the presence of both in our cellular population likely 
reflects the detailed growth and acclimatization conditions.  For consistency,
we focus here only on positively phototactic cells.
For them, the residual torque provides information on the pitch and amplitude of helical 
trajectories of unstimulated cells.

\begin{figure}[t]
\includegraphics[width=0.80\columnwidth]{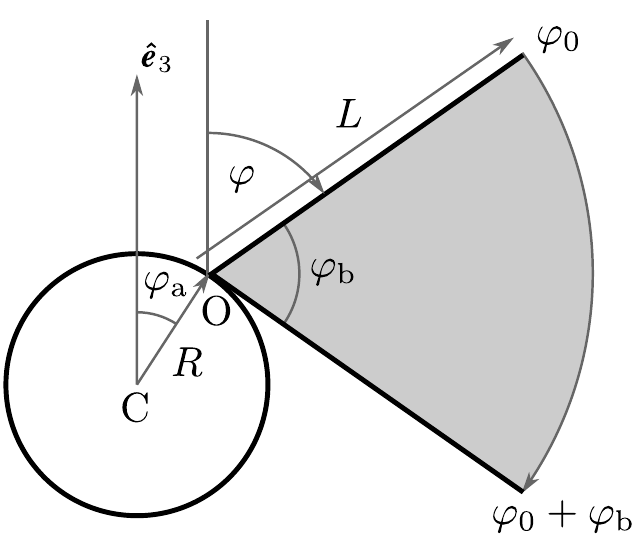}
\caption{The pivoting-rod model of the power stroke.}
\label{fig7}
\end{figure}

\subsection{Heuristic model of flagellar beating}
\label{heuristic}

Below we extend the quantification of flagellar beating to 
a transient photoresponse like that in Fig.~\ref{fig4}, with the 
goal of inferring the angular velocity 
$\omega_1$ around $\hat{\bf e}_1$
that a freely-swimming cell would experience, and which leads to a phototurn.  
The constant of proportionality
between the torque and the angular velocity is an effective rotational drag coefficient
that can be viewed as one of a small number of parameters of the overall 
phototaxis problem.  Our immediate goal is to develop an estimate of
this constant to provide a consistency check on the final theory in its
comparison to experiment.

\begin{figure}[t]
\includegraphics[width=0.98\columnwidth]{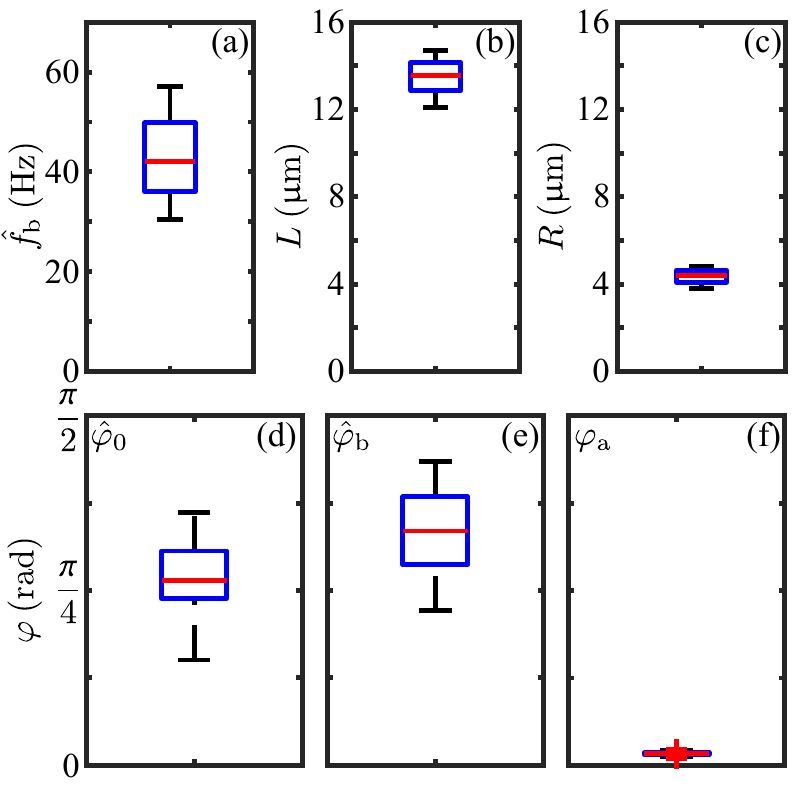}
\caption{Distributions of geometric quantities for flagellar beats, 
computed from $n=24$ movies: (a)  
instantaneous flagellar beat frequency $\hat{f}_\mathrm{b}$ (b) flagellar length $L$, (c) cell-body radius $R$, 
(d-f) per-beat flagella-line initial angle $\hat{\varphi}_0$, sweep angle $\hat{\varphi}_b$ and anchor angle $\varphi_a$.}
\label{fig8}
\end{figure}

A simple model of the power stroke shown in  Fig.~\ref{fig7} can be used 
to understand the peak values $\hat{\mathcal F}^*$ and $\hat{\mathcal T}^*$:
a straight flagellum attached at angle $\varphi_a$ to a spherical body
of radius $R$, whose beat angle 
$\varphi(t)$ sweeps from $\hat{\varphi}_0$ to $\hat{\varphi}_0+\hat{\varphi}_b$.
Table \ref{table1} and Fig.~\ref{fig8} summarize
data on these geometric quantities.
Relative to the center $\mathrm{C}$ a point at arclength $\lambda\in [0,L]$ is 
at position
\begin{equation}
{\bf r}(\lambda,\varphi) = {\bf R}_a+\lambda \hat{\bf t}(\varphi),
\label{rod_param}
\end{equation} 
where ${\bf R}_a=R\left[\sin{\varphi_a}\hat{\bf e}_x + \cos{\varphi_a}\hat{\bf e}_y\right]$
is the vector $\mathrm{CO}$ and
$\hat{\bf t}(\varphi)=\sin{\varphi}\,\hat{\bf e}_x + \cos{\varphi}\,\hat{\bf e}_y$
is the unit tangent.  The velocity of a point on the filament is 
${\bf r}_t(\lambda,\varphi)=-\lambda\dot\varphi \hat{\bf n}$.

The integrated force of the pivoting rod in Fig.~\ref{fig5} is
\begin{equation}
    \bm{{\cal F}}=\frac{1}{2}\zeta_\perp L^2\dot{\varphi}\hat{\bf n},
    \label{oneflag}
\end{equation}
Considering the sinusoidal variation of the quantities 
in Figs. \ref{fig6}(a,c), we estimate $\varphi$ by the lowest mode that
has vanishing speed at beginning and end of the power stroke,
\begin{equation}
    \varphi(t)=\hat{\varphi}_0+\frac{1}{2}\hat{\varphi}_b\left[1-\cos\left(\pi t/\eta T_b\right)\right],
    \label{phi_param}
\end{equation}
for $t\in [0,\eta T_b]$, where $T_b$ is the full beat period and $\eta\simeq 0.7$ is the fraction
of the period occupied by the power stroke. Using the data in Table \ref{table1}, and the fact 
that the maximum projected force occurs very close to the time when $\varphi=\pi/2$, we obtain
the estimate 
\begin{equation}
\hat{\cal F}^*\sim \frac{\pi^2\mu L^2 f_b\hat{\varphi}_b}{\eta c_\perp}\sim 22\,{\rm pN},
\label{fmax}
\end{equation}
which is $\sim\!1$ s.d. above the experimental mean.
While it is not surprising that the pivoting rod model overestimates the propulsive force relative
to the actual undulating flagellum, the fact that this overestimate
is small indicates that the essential physics is contained in \eqref{fmax}.

Further heuristic insight into the flagellar forces produced can be gained by estimating
the resultant motion of the cell body, assumed to be a sphere of radius $R$.  This requires
incorporating the drag of the body and that due to the flagella themselves.
A full treatment of this problem requires going beyond RFT to account for the effect of
flows due to the moving body on the flagellar and vice versa.  In the spirit of the rod model, 
considerable insight can be gained in the limit of very long flagella, where the fluid flow
is just a uniform translational velocity ${\bf u}(t)$ and 
the velocity of a point on the rod is
\begin{equation}
    {\bf r}_t(\lambda,\varphi)=-\lambda\dot\varphi \hat{\bf n}+{\bf u}.
    \label{generalmotion}
\end{equation}

Symmetry dictates that 
the net force from the downward sweeps of two mirror-image flagella is along $\hat{\bf e}_3$, as is
the translational velocity ${\bf u}=u\hat{\bf e}_y$
of the cell body. Adding mirror-image copies of the force \eqref{oneflag} and
the drag force on the body
$-\zeta u\hat{\bf e}_y$, where
$\zeta=6\pi\mu R$ is the Stokes drag coefficient for a sphere of radius $R$, the condition that
the total force vanish yields
\begin{equation}
    u(t)=v \frac{\sin\varphi\sin\left(\pi t/\eta T_b\right)}{1+d_\perp \sin^2\varphi
    +d_\parallel \cos^2\varphi},
    \label{Uestimate}
\end{equation}
where 
$d_\perp=2\zeta_\perp L/\zeta=4L/3R c_\perp$ and $d_\parallel=2L/3R c_\parallel$.  The speed
$v$ is given by the maximum force \eqref{fmax} as
\begin{equation}
    v=\frac{2\hat{\mathcal F}^*}{\zeta}=\frac{\pi f_b\phi_b L^2}{3\eta c_\perp R},
       \label{Uestimate2}
\end{equation}
and is independent of the viscosity 
$\upmu$, as it arises from a balance between the two drag-induced forces of flagellar propulsion and
drag on the spherical body, the latter ignoring the contribution in the denominator of \eqref{Uestimate} from flagellar drag. 
For typical 
parameters, $d_\perp\approx 0.8$, and the denominator is $\approx\! 1.8$ when $u$ is
maximized (at $\varphi=\pi/2$), while $v\sim 540\,$\textmu m/s.  Thus,
the peak swimming speed during the power stroke would be $u^*\sim v/1.8 \sim 300\,$\textmu m/s,
consistent with measurements \cite{Guasto2010}, 
which also show that over a complete cycle, including the recovery stroke, the 
mean speed $\bar{u}\sim u^*/4$.  We infer that 
$\bar{u}\sim 75\,\upmu$m/s, consistent with observations \cite{Guasto2010,Leptos2009}.

We now use the rod model to estimate the maximum torque $\hat{\mathcal T}^*$ produced
on an immobilized cell, to compare with the RFT calculation from
the experimental waveforms.
As in \eqref{generalmotion}, the force density on the moving filament is 
${\bf f}=\zeta_\perp \lambda \dot{\varphi}\hat{\bf n}$, the torque density is
$\left({\bf R}+\lambda\hat{\bf t}\right)\times {\bf f}$, and the integrated torque component
along $\hat{\bf e}_1$ is
\begin{equation}
    \hat{\mathcal T}=\frac{1}{2}\zeta_\perp\dot{\varphi}R^3
    \left(\ell^2\cos\left(\varphi-\varphi_a\right)+\frac{2}{3}\ell^3\right).
    \label{torque_eqn}
\end{equation}
where $\ell=L/R$ and $\varphi$ is again given by \eqref{phi_param}. The two terms 
in \eqref{torque_eqn}, scaling as $RL^2$ and $L^3$, arise from
the distance offset from the cell body and the integration along the flagellar length, 
respectively.

Examining this
function numerically we find that its peak occurs approximately
midway through the power stroke, where $\varphi-\varphi_a\simeq \pi/3$, leading
to the estimate
\begin{equation}
    \hat{\mathcal T}^*=\frac{\pi^2\mu f_b\hat{\varphi}_b R^3}{\eta c_\perp}
    \left(\frac{1}{2}\ell^2+\frac{2}{3}\ell^3\right)\sim 250\,{\rm pN}\cdot \upmu{\rm m}.
\end{equation}
and, for average torque,
\begin{equation}
    \langle\hat{\mathcal T}\rangle\simeq\frac{2}{\pi}{\mathcal T}^*\sim 160\,{\rm pN}\cdot \upmu{\rm m}.
    \label{average_torque}
\end{equation}
Here again these estimates are slightly more than $1$ s.d. above the experimental 
value, giving 
further evidence that the rod model is a useful device to understand the scale of 
forces and torques of beating flagella.

The essential feature of \eqref{Uestimate} is 
an effective translational drag coefficient $\tilde{\zeta}$ that is larger than that of the sphere 
due to the presence
of the very beating flagella that cause the motion.  For flagella oriented at $\varphi=\pi/2$, 
$\tilde{\zeta}=\zeta+2\zeta_\perp L$,
a form that reflects the extra contribution from transverse drag on the two flagella.
We now consider the analogous rotational problem and estimate an
effective rotational drag coefficient $\tilde \zeta_r$ in terms of the bare rotational
drag $\zeta_r=8\pi\mu R^3$ for a sphere. 
If we set in rotational motion at angular speed ${\bm\Omega}$ a sphere with two flagella 
attached at angles $\pm\varphi_a$, 
the velocity of a rod segment at $\lambda$
is ${\bm \Omega}\times {\bf r}$ and the calculation of the
hydrodynamic force and torque proceeds as before,
yielding  
\begin{align}
    \frac{\tilde{\zeta_r}}{\zeta_r}=
    1&+\frac{\ell\cos^2(\varphi-\varphi_a)+\ell^2\cos(\varphi-\varphi_a)+\ell^3/3}{c_\perp}\nonumber \\
    &+\frac{\ell \sin^2(\varphi-\varphi_a)}{2c_\parallel}.
    \label{dragratio}
\end{align}
For typical parameters, $\zeta_r\sim 2.1\,$pN$\cdot\upmu$m$\cdot$s, and the added drag of the flagella is significant; the ratio in \eqref{dragratio} varies from 
$4.8-3$ as $\varphi$ varies from $\hat{\varphi}_0$ to 
$\hat{\varphi}_0+\hat{\varphi}_b$.  At the approximate peak of the power stroke we find 
$\tilde{\zeta}_r\simeq 9\,$pN$\cdot\upmu$m$\cdot$s, a value we use in further estimates below.

The effective rotational drag coefficient can be used to estimate the
unstimulated angular speed due to the small torque imbalance
noted below \eqref{residual_torque} for the {\it cis}-dominant subpopulation,
\begin{equation}
    \hat{\omega}_1\sim\frac{\langle \hat{\mathcal T}\rangle^{\rm cis}}{\tilde{\zeta}_r}
\simeq 1\,{\rm s}^{-1},
\label{hatomega1}
\end{equation}
which can be compared to the angular speed
$\vert\omega_3\vert\simeq 10\,$s$^{-1}$ of spinning around the primary axis.  
In Sec. \ref{sec:turndynamics} we show that
the small ratio $\hat{\omega}_1/\vert\omega_3\vert\sim 0.1$ implies that the 
helices are nearly straight.

\subsection{Adaptive dynamics}
\label{sec:adaptive}

\begin{figure}[t]
\includegraphics[width=0.98\columnwidth]{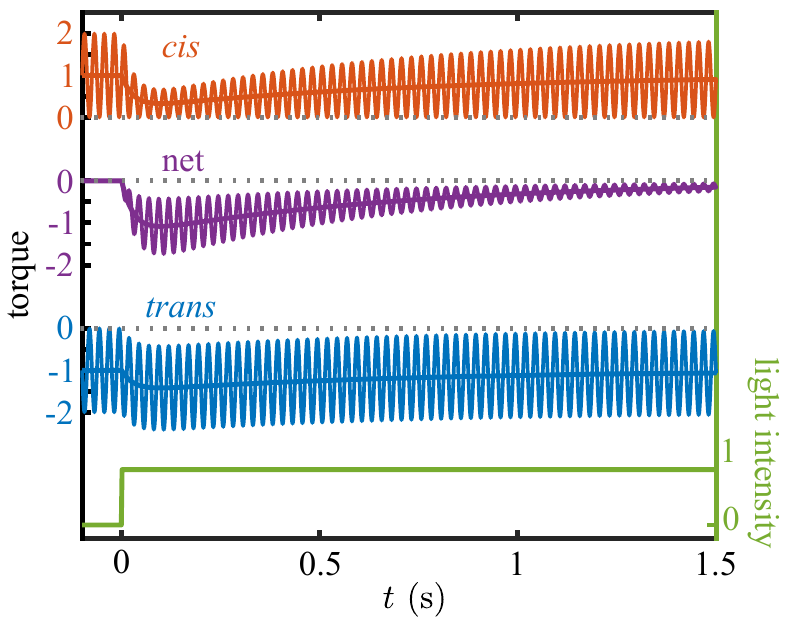}
\caption{Schematic of flagellar photoresponse.  A step-up
in light at $t=0$ (green) leads
to a biphasic decrease in the mean value and oscillation amplitude of the {\it cis} phototorque (red), and a 
biphasic increase in the magnitude of the mean
value of the {\it trans} phototorque (blue). The net torque (purple), the signed sum of the two contributions, 
has a biphasic response in both the 
oscillation amplitude and its running mean value.}
\label{fig9}
\end{figure}

The results of the previous section constitute
a quantitative understanding of the phototactic torques
produced {\it within} a given flagellar beat, which
typically lasts $20-25\,$ms.  
As mentioned previously, the timescale for
the full photoresponse associated with a 
change in light levels falling on the eyespot is
considerably longer, on the order of $0.5\,$s.
This separation of timescales is illustrated in 
Fig.~\ref{fig9}, where we have schematically
shown the time-resolved, oscillating phototactic torque of each of the two flagella, the signed sum,
and its running average.  It is precisely because of the
separation of timescales between the rapid beating and
both the slow response and the slow phototurns that a theory developed in terms of the beat-averaged torques is justified.

\begin{figure*}[t]
\includegraphics[width=1.98\columnwidth]{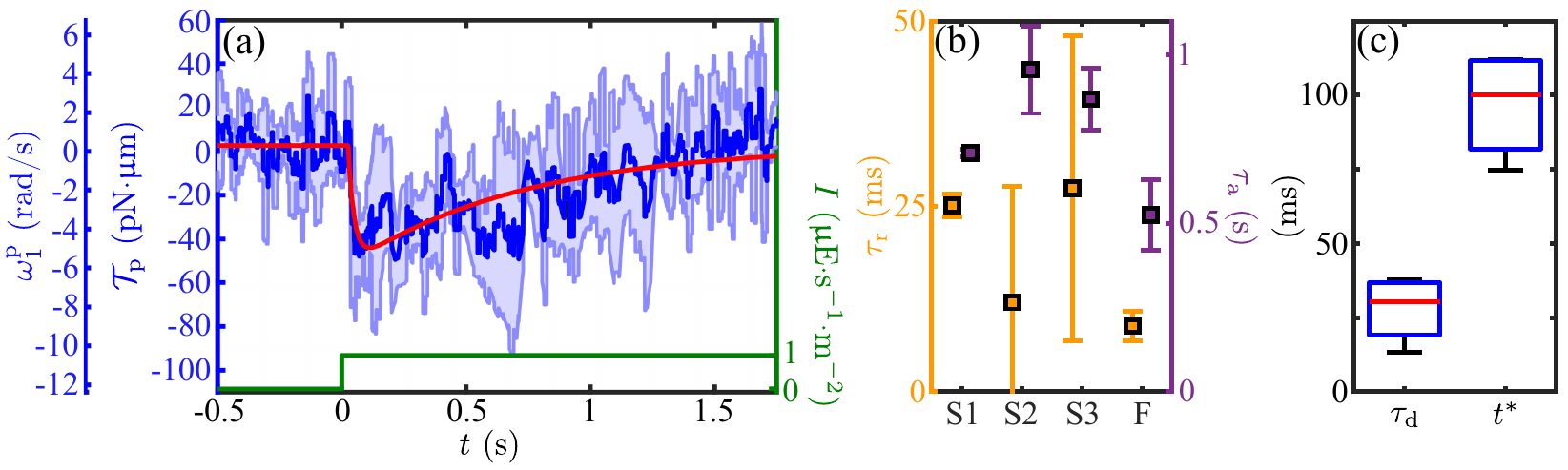}
\caption{Dynamics of the flagellar photoresponse.
(a) The mean (dark blue) and standard deviation (light-blue) 
of the proxy torque during a step-up stimulus for one cell ($n_\mathrm{tech}$ = 4) 
fitted to \eqref{eq:p_sln} (red line). (b) Inset showing fitted ($\tau_r$, $\tau_a$) pairs 
for $n_\mathrm{cells}$ = 4 upon step-up stimulation. (c) As in (b) but for the delay time $\tau_d$ and time
of maximum photoresponse amplitude $t^*$.}
\label{fig10}
\end{figure*}

In the following, we measure phototorques relative to
the unstimulated state of the cell, and define the 
two (signed) {\it beat-averaged} quantities
\begin{equation}
    \delta{\mathcal T}_{\rm cis}=\langle {\mathcal T}_{\rm cis}\rangle -\langle \hat{\mathcal T}_{\rm cis}\rangle, \ \ \ \ 
    \delta{\mathcal T}_{\rm trans}=\langle {\mathcal T}_{\rm trans}\rangle -\langle \hat{\mathcal T}_{\rm trans}\rangle,
\end{equation}
and their sum, the net beat-averaged phototactic torque 
\begin{equation}
    {\mathcal T}_{\rm p}=\delta{\mathcal T}_{\rm cis}+\delta{\mathcal T}_{\rm trans}.
\end{equation}
${\mathcal T}_{\rm p}>0$
when the {\it cis} flagellum beats more strongly and ${\mathcal T}_{\rm p}<0$ when
the {\it trans} flagellum does.  Our strategy is to determine ${\mathcal T}_{\rm p}$
from experiment on pipette-held cells and to estimate the resulting angular speed $\omega_1^{\rm p}$ using
$\tilde{\zeta}_r$.

The scale of net torques expected during a transient
photoresponse can be estimated from the pivoting-rod 
model.  
From step-up experiments such as that shown in Fig.~\ref{fig4}, we observe that there are
two sweep angles $\varphi_b^{\rm cis,trans}$ whose difference
$\Delta\varphi_b^*=\max\{\varphi_b^{\rm cis}-\varphi_b^{\rm trans}\}$ can be used in \eqref{average_torque} to obtain 
${\mathcal T}_{\rm p}^*$,
the maximum value of the beat-averaged sum (corresponding to
the most negative value of the purple running mean in 
Fig. \ref{fig9}).  Averaging over
$8$ photoresponse videos, we find $\Delta\varphi_b^*\sim -\pi/14$, which 
yields the estimate
\begin{equation}
    {\mathcal T}_{\rm p}^* \sim -34\,{\rm pN}\cdot \upmu{\rm m}.
    \label{torque_estimate}
\end{equation}
From the effective drag coefficient the corresponding peak
angular speed in such a photoresponse is 
\begin{equation}
    \omega_1^{*} \sim \frac{{\cal T}_{\rm p}^*}{\tilde{\zeta}_r}\sim -4\,{\rm s}^{-1}.
    \label{omega1star_est}
\end{equation}
To put this in perspective, consider the photoalignment 
of an alga swimming initially perpendicular to a light source.  If sustained continuously,
complete alignment would occur in a time $(\pi/2)/\omega_1^* \sim 0.4\,$s, whereas our observations
suggest a longer timescale of $\sim 2\,$s.  This will be shown to follow from the
variability of $\omega_1$ during the trajectory in accord with an adaptive dynamics.

While the estimate in \eqref{torque_estimate} gives a guide to scale
of the torques responsible for phototurns, we may calculate 
them directly within RFT from flagellar beating asymmetries 
in the same manner as in the unstimulated case.
Figure \ref{fig10} shows the response of a single cell to a step-up in illumination 
(of which Fig.~\ref{fig4} is a snapshot),
in which the results are presented both in terms of ${\mathcal T}_{\rm p}$ and
the estimated $\omega_1^{\rm p}$.  To obtain these data, the oscillating time series of {\it cis} and {\it trans} 
torques were processed to obtain beat-averaged values whose sum yields the
running average, as in Fig.~\ref{fig9}. The overall response is $<0$, indicating 
that the {\it trans} flagellum dominates, and the peak value averaged over multiple cells 
of $ - 37\pm 12\,$pN$\cdot$nm is 
consistent with the estimate in \eqref{torque_estimate}. 
The biphasic response, with a 
rapid increase followed by a slow return to zero, 
is the same form observed in 
{\it Volvox} \cite{Drescher2010} and {\it Gonium} \cite{GoniumPRE}.

\begin{figure}[b]
\includegraphics[width=0.98\columnwidth]{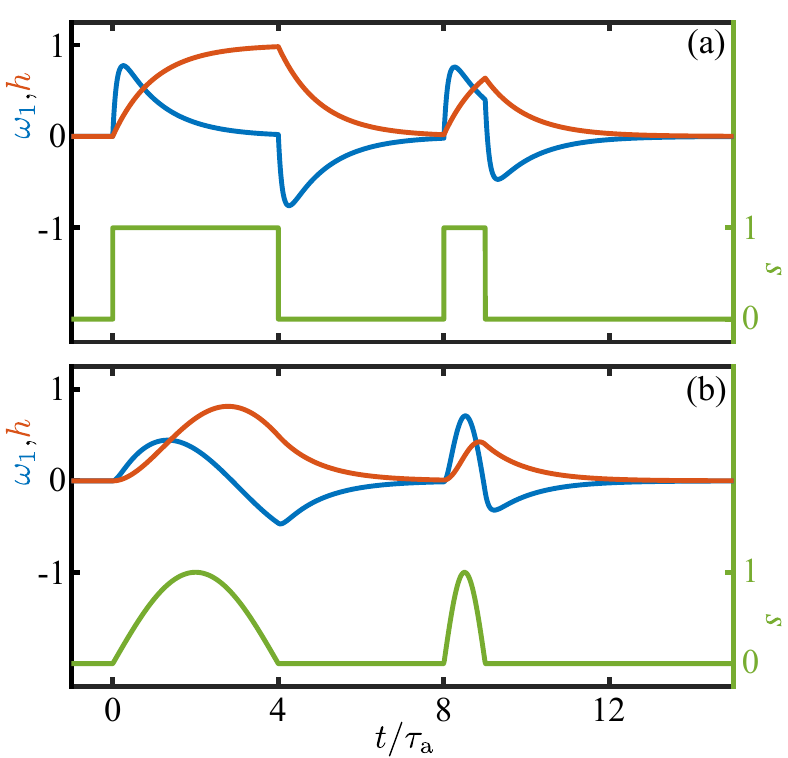}
\caption{Dynamics of the adaptive model.  (a) Response of the variables $p$ and $h$ 
to a square pulse of stimulus (green), for $\rho=0.1$. (b) Response to rectified sinusoids.
}
\label{fig11}
\end{figure}

We now argue that the adaptive model \eqref{eq:model} used for those cases can 
be recast as an 
evolution equation for the
angular speed itself, setting $p=\omega_1$ and $s=g I$, with $I$ the light intensity and 
$g$ a proportionality constant,%
\begin{subequations}
\begin{align}
\tau_r\dot{\omega_1}&=s-h - \omega_1 \\
\tau_a\dot{h}&=s-h.
\end{align}
\label{eq:model_repeated}%
\end{subequations}
For constant $s$ the system \eqref{eq:model_repeated} has the fixed point $(\omega_1^*,h^*)=(0,s)$.
If $\omega_1=h=s=0$ for $t\le 0$, followed by $s=s_0$ for $t>0$, then
\begin{subequations}
\begin{align}
	\omega_1(t)&=\frac{s_0}{1-\rho}\left(e^{-t/\tau_a}-e^{-t/\tau_r}\right) \label{eq:p_sln} \\
	h(t)&=s_0\left(1-e^{-t/\tau_a}\right),
\end{align}
\label{model_response}%
\end{subequations}
where $\rho=\tau_r/\tau_a$. The result \eqref{eq:p_sln}, illustrated in Fig.~\ref{fig11}(a) for the
case $\rho=0.1$ and a square pulse of duration long compared to $\tau_a$, 
shows clearly the biphasic response of the data in Fig.~\ref{fig10}. 
This behavior is like two coupled
capacitors charging and discharging one another, particularly in the limit $\rho\ll 1$.
At early times, $h$ remains small and $\omega_1$ relaxes toward $s_0$ with the rapid
timescale $\tau_r$.  Later, when $t\sim \tau_a$, $h$ relaxes toward $s_0$, and
$\omega_1$ relaxes instead toward zero, completing the pulse.

After a step up, $h$ has relaxed to $s_0$, and if $s$
is then stepped down to zero, $\omega_1$ rapidly tends toward $-s_0$, then
later reverses its negative growth and returns to zero.  If, as in Fig.~\ref{fig11}, the pulse width is much
larger than $\tau_a$, the step down response is simply the negative of the step-up response.
For smaller step duration, the step-down response is still negative, but is not a
mirror image of the
step-up dynamics.
Taking $s_0$ to be positive, this antisymmetric response implies that 
as the eyespot rotates into the light there is a step-up response with $\omega_1>0$, corresponding to
transient {\it cis} flagellar dominance, and when the eyespot rotates out of the light then
$\omega_1 <0$, associated with {\it trans} flagellum dominance.  This is precisely
the dynamics shown in Fig.~\ref{fig1} that allows monotonic turning toward the light
as the cell body rotates.

Note that the adaptive dynamics coupling $\omega_1$, $s$, and $h$ is left unchanged by the
simultaneous change of signs $\omega_1\to -\omega_1$, $h\to -h$, and $s\to -s$.  This 
symmetry allows us
to address positive and negative phototaxis in a single model, for if
a step-up in light activates a transient dominant {\it trans} flagellum response 
in the cell orientation
of Fig.~\ref{fig2}, with $\omega_1 < 0$, we need only take $s<0$.

Since the model \eqref{eq:model_repeated} is constructed so that $\omega_1$ is forced by the 
signal $s$, the opposite-sign response to step-up and step-down signals is not an obvious
feature.  Yet, in the standard manner of coupled first-order ODEs, the hidden variable
$h$ can be eliminated, yielding a single, second-order equation for $\omega_1$.  It can be
cast in the simple form
\begin{equation}
    \left(\tau_r\frac{d}{dt}+1\right)\left(\tau_a\frac{d}{dt}+1\right)\omega_1=\tau_a\dot{s},
    \label{model_single}
\end{equation}
which is explicitly forced by the derivative of the signal, thus driven 
oppositely during and and down steps.

Previous studies \cite{Josef2006} found a
measurable time delay $\tau_d$ between the signal and the response that, in the language of
the adaptive model, is additive with the intrinsic offset determined by the time scales $\tau_r$ and
$\tau_a$.  This can be captured by expressing the signal in \eqref{eq:model_repeated} as $s(t-\tau_d)$.
The maximum amplitude of $\omega_1(t)$ then occurs at the time
\begin{equation}
t^* = \frac{\tau_r}{1-\rho}\ln(1/\rho)+\tau_d,
\label{eq:t_max}
\end{equation}
at which point the amplitude is $s_0 A(\rho)$, where
\begin{equation}
    A(\rho)=\frac{\rho^{\rho/1-\rho}-\rho^{1/1-\rho}}{1-\rho}.
    \label{peakamp}
\end{equation} 
A fit to the step-response data yields
$\tau_d=28\pm 11\,$ms and $t^*=97\pm 18\,$ms.
This delay between stimulus and maximum response has a geometric interpretation.
The angle through which the eyespot rotates
in the time $t^*$ is $\vert\omega_3\vert t^*= (0.28\pm 0.05)\pi$, which is very nearly 
the angular shift $\kappa = \pi/4$ of the eyespot location from 
the ($\hat{\bf e}_2$-$\hat{\bf e}_3$) flagellar beat plane (Fig.~\ref{fig1}).
Since $\omega_3<0$, 
the eyespot {\it leads} the flagellar plane and thus $t^*$ is the time needed
for the beat plane to align with the light.  In this configuration,
rotations around 
$\hat{\bf e}_1$ are
most effective \cite{Schaller1997,Josef2006}.

Since the function $A(\rho)$ decreases monotonically from unity at $\rho=0$, we identify 
the maximum angular speed $\omega_1^*$ attainable for a given a stimulus as $s_0$.  With $s_0=g I$,
we can remove $g$ from the problem by instead viewing $\omega_1^*$ as the fundamental
parameter, setting
\begin{equation}
    s(t)=\omega_1^*(I).
    \label{step_single}
\end{equation}
As indicated in \eqref{step_single}, there is surely a dependence of $\omega_1^*$ on the light intensity,
not least because the rotational speed will have a clear upper bound associated with the limit in which
the subdominant flagellum ceases beating completely during the transient photoresponse.

\begin{figure}[t]
\includegraphics[width=0.98\columnwidth]{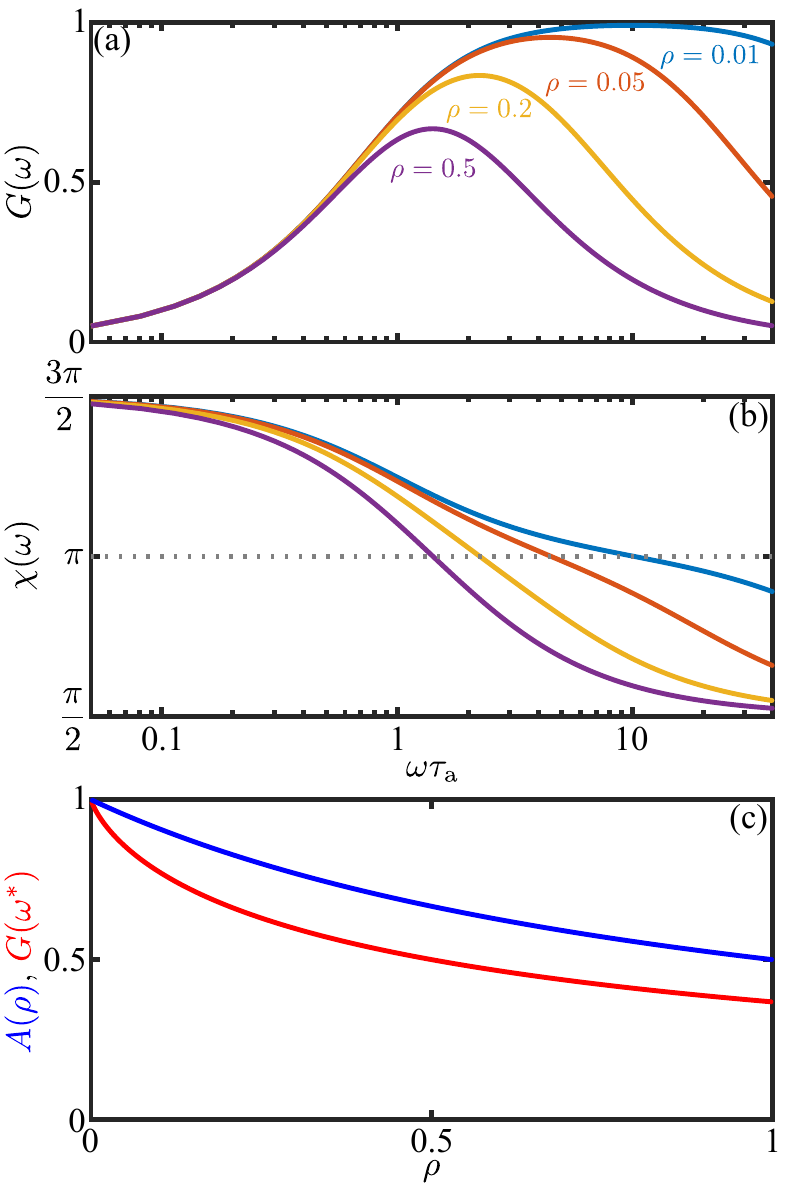}
\caption{Dynamics of the adaptive model.  (a) Gain \eqref{gainphase_a} for various 
values of $\rho=\tau_r/\tau_a$. (b)  As in (a), but for the phase shift $\chi$ \eqref{gainphase_b},
with $\tau_d=0$. 
(b) Comparison of 
peak amplitude for step-up and oscillatory forcing as a function of $\rho$.
}
\label{fig12}
\end{figure}

Turning now to the oscillating light signals experienced by freely-rotating cells, the 
directionality of the eyespot implies that the signal will be a
{\it half-wave rectified} sinusoid (HWRS).  Figure \ref{fig11}(b) shows the response
of $\omega_1$ to two single half-period signals of this type.  Compared to the
square pulses of equal duration and maximum [Fig.~\ref{fig11}(a)] the maximum response
amplitude is reduced due to the lower mean value and slower rise of the signal.  
The frequency response of the adaptive model is most easily deduced from
\eqref{model_single}, and if $s(t)=\tilde{s}e^{i\omega t}$, then
there will be a proportionate amplitude $\tilde{\omega}_1$.  We define the response
$\mathscr{R}(\omega)= \tilde{\omega}_1/\tilde{s}$, gain  
$G(\omega)=|\mathscr{R}(\omega)|$ and phase shift
$\chi=\tan^{-1}[{\rm Im}(\mathscr{R})/{\rm Re}(\mathscr{R})]$.  These are
\begin{subequations}
\begin{align}
 G(\omega)&= \frac{\alpha}{\sqrt{\left(1+\alpha^2\right)
\left(1+\rho^2\alpha^2\right)}} \label{gainphase_a}\\
   \chi&=\pi+\chi_0-\delta, \ \ \ \ \chi_0=\tan^{-1}\left(\frac{1-\rho\alpha^2}{\left(1+\rho\right)\alpha} \right),
   \label{gainphase_b}
   \end{align}
   \label{gainphase}%
   \end{subequations}
where $\alpha=\omega\tau_a$, $\delta=\omega\tau_d$ and the additive term of $\pi$ in the phase represents the 
sign of the overall response.
Figures \ref{fig12}(a,b) show these quantities as a function of the stimulus frequency 
$\omega$ for
various values of $\rho$.  The peak frequency is at $\alpha^*=1/\sqrt{\rho}$, or
$\omega^*=1/\sqrt{\tau_r\tau_a}$,
at which $G(\omega^*)=1/(1+\rho)$ and $\chi=\pi-\omega^*\tau_d$.  
Fig.~\ref{fig12}(a) shows that the peak is 
sharp for large $\rho$ and becomes much broader 
as $\rho\to 0$. 
The peak amplitude decays in a manner similar to \eqref{peakamp} for a step response
[Fig.~\ref{fig12}(c)].

\begin{figure*}[t]
\includegraphics[width=1.98\columnwidth]{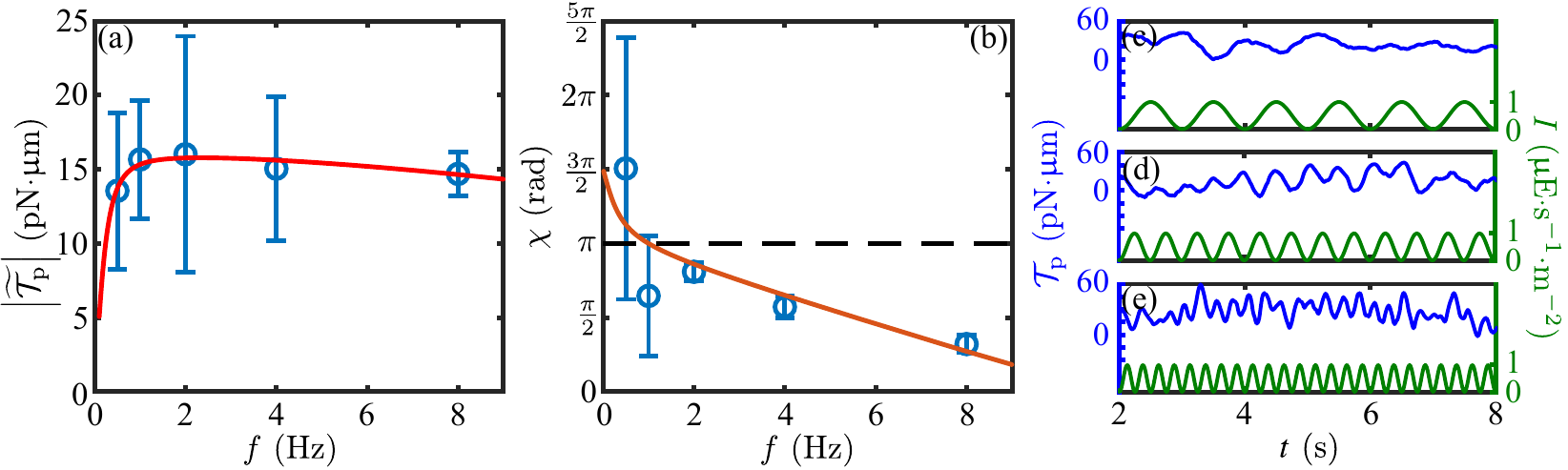}
\caption{Frequency response of immobilized cells.
(a) Measured beat-average phototactic torque determined from RFT 
(for positive phototaxis) at five stimulus frequencies 
(0.5, 1, 2, 4 and 8 Hz) for $n_\mathrm{cells} = 3$ (blue) fitted to 
\eqref{gainphase_a} (red line). (b) As in (a), but for the phase $\chi$ 
of the response, fitted to \eqref{gainphase_b}. The photoresponse 
shown (in blue) for three different stimulus frequencies (in green): 
(c) $1$ Hz, (d) $2$ Hz, and (e) $4$ Hz.}
\label{fig13}
\end{figure*}

The peaked response function amplitude \eqref{gainphase_a} and phase shift \eqref{gainphase_b} are
qualitatively similar to those obtained experimentally by Josef, {\it et al.} \cite{Josef2006},
who analyzed separately the {\it cis} and {\it trans} responses and found distinct peak
frequencies for the two, and investigated the applicability of more complex 
frequency-dependent response functions than those in \eqref{gainphase}. 
In the spirit of the analysis presented here we do not
pursue such detailed descriptions of the flagellar responses, but it 
would be straightforward to incorporate them as we discuss in Sec. \ref{photosteering}.

Using the same protocol as for the step function response in Fig.~\ref{fig10}, we 
measured the frequency dependent photoresponse by subjecting
cells to an oscillating light intensity at five distinct frequencies, analyzing the
transient waveforms using RFT and determining the beat-average torque magnitude.  
The results of this
study (Fig.~\ref{fig13}), were fit to the form \eqref{gainphase_a}, from which
we obtained the time constants $\tau_r=0.009\pm 0.002\,$s and $\tau_a=0.52\pm 0.10\,$s, and thus $\rho=0.02$.
This strong separation between response and adaptation time scales is consistent with that
seen under the step response (Fig.~\ref{fig10}(a,b)) and leads to the broad peak of 
the frequency response curve. The peak frequency ($\simeq 2\,$ Hz) is in very close agreement with recent direct measurements of rotation frequency about the cell body axis on free-swimming cells \cite{Choudhary}. 
The phase data shown in Fig.~\ref{fig13}(b)
are well-described by the adaptive model with the parameters determined from the fit to the 
amplitude, with a time delay of
$\tau_d=38\pm 5\,$ms, a value that is consistent with that obtained from the
step response.  Note that for frequencies $\omega$ near $\omega^*$ and for
$\rho \ll 1$, the phase has the simple form
\begin{equation}
    \chi\simeq \pi -2\tau_r\left(\omega-\omega^*\right)-\omega^*\tau_d + \cdots.
\end{equation}
This result shows that while negative detuning from $\omega^*$ by itself increases the
phase above $\pi$, the time delay can be a more significant contribution, leading to
$\chi<\pi$.  Such is indeed the case in Fig.~\ref{fig13}, where the peak frequency is $\simeq 2\,$Hz,
but $\vert\omega_3\vert\tau_d\simeq 0.13\pi$ and $\chi(\omega^*)<\pi$.

\section{Dynamics of phototactic turns}
\label{sec:turndynamics}

\subsection{Helices, flagellar dominance, and eyespot shading}

We now consider the larger length scales associated with
the swimming trajectory of cells, and note
the convention that rotation around an axis ${\bf e}_i$ is taken to have a {\it positive} angular 
velocity $\omega_i$ if the rotation is {\it clockwise} when viewed along the direction 
that ${\bf e}_i$ points.
{\it Chlamydomonas} 
spins about ${\bf e}_3$ with an angular
velocity $\omega_3<0$, and we define the positive frequency $f_r=-\omega_3/2\pi$.  
Its helical trajectories arise from an additional 
angular velocity $\hat{\omega}_1{\bf \hat{e}}_1$, and    
we assume that $\omega_1$, $\omega_3$, 
and the translational speed $u$ along ${\bf \hat{e}}_3$ are
sufficient to define the trajectories, without invoking 
an angular velocity $\omega_2$.

The natural description of swimming trajectories is through the 
Euler angles ($\phi,\theta,\psi$) that define its orientation.  In the standard 
convention \cite{Goldstein_mechanics}, 
their time evolution 
is given by angular velocities ($\omega_1,\omega_2,\omega_3$) as follows,%
\begin{subequations}
\begin{align}
    \omega_1&=\dot{\phi}\sin\theta \sin\psi+\dot{\theta}\cos\psi \\
    \omega_2&=\dot{\phi}\sin\theta \cos\psi-\dot{\theta}\sin\psi \\
    \omega_3&=\dot{\phi}\cos\theta+\dot{\psi}.%
\end{align}
\label{Eulerangledynamics}%
\end{subequations}
The transformation from the body frame ${\bf x}$ to the laboratory frame ${\bf x}'$
is ${\bf x}'={\bf A}\cdot {\bf x}$ and the reverse transformation is via
${\bf x}=\tilde{\bf A}\cdot {\bf x}'$, where $\tilde{\bf A}={\bf A}^{-1}={\bf A}^T$, with
\begin{equation}
\tilde{\bf A}=
\begin{pmatrix}
{\rm c}\psi\, {\rm c}\phi - {\rm c}\theta\, {\rm s}\phi\, {\rm s}\psi& -{\rm s}\psi\, {\rm c}\phi 
-{\rm c}\theta\, {\rm s}\phi\, {\rm c}\psi&{\rm s}\theta\, {\rm s}\phi \\ 
{\rm c}\psi\, {\rm s}\phi+ {\rm c}\theta\, {\rm c}\phi\, {\rm s}\psi&-{\rm s}\psi\, {\rm s}\phi 
+ {\rm c}\theta\, {\rm c}\phi\, {\rm c}\psi& -{\rm s}\theta\, {\rm c}\phi  \\
{\rm s}\theta\, {\rm s}\psi&{\rm s}\theta\, {\rm c}\psi& {\rm c}\theta 
\end{pmatrix},
\label{tildeA}
\end{equation} 
and we have adopted the shorthand $c\psi=\cos\psi$, etc.  

The connection between helical swimming trajectories and the angular velocities $\omega_i$ has 
been made by Crenshaw \cite{Crenshaw1,Crenshaw2,Crenshaw3} by first 
postulating helical motion and then finding consistent angular velocities. 
We use a more direct approach, starting from the Euler angle dynamics \eqref{Eulerangledynamics}. 
If there is motion along a helix, and $\hat{\omega}_1$ and $\omega_3$ 
are nonzero and constant, then apart from the degenerate case of
orientation purely along $\hat{\bf e}_z$, where ``gimbal locking" occurs, we must have
$\dot{\phi}={\rm constant}$, $\dot\theta=0$, and $\dot{\psi}=0$.  If we thus set 
$\dot{\phi}=-\gamma$ (the sign choice taken for later convenience), 
$\theta=\theta_0$ (with $-\pi/2\le \theta_0 \le \pi/2)$, then 
a solution requires $\psi=\pi/2$ and the
primary body axis is
\begin{equation}
    {\bf \hat{e}}_3=-\sin\theta_0\left[\sin\gamma t\,{\bf \hat{e}}_x 
    +\cos\gamma t\,{\bf \hat{e}}_y\right]
    +\cos\theta_0\, {\bf\hat {e}}_z.
    \label{tangent_helix}
\end{equation}
If the organism swims along the positive ${\bf \hat{e}}_3$ direction at speed $u$,
then $\hat{\bf e}_3$ is the tangent vector $\hat{\bf t}$ to its trajectory and we can integrate 
\eqref{tangent_helix} using ${\bf \hat{t}}=(1/u){\bf r}_t$ to obtain 
\begin{equation}
    {\bf r}(t)=\frac{u\sin\theta_0}{\gamma}\left[\cos\gamma t\,{\bf \hat{e}}_x
    -\sin\gamma t\,{\bf \hat{e}}_y\right]+ut\cos\theta_0\,{\bf \hat{e}}_z,
    \label{helix_general}
\end{equation}
which is a helix of radius $R_h$ and pitch $P_h$ given by
\begin{equation}
    R_h= \frac{u\left\vert\sin\theta_0\right\vert}{\left\vert\gamma\right\vert},  \ \ \ \ 
    P_h=\frac{2\pi u\cos\theta_0}{\vert\gamma\vert}.
\end{equation}

With the parameters $(\gamma,\theta_0)$ taking either positive or negative values, there
are four sign choices: $(+,+)$, $(+,-)$, $(-,+)$, and $(-,-)$.  Since the $z$ coordinate
in the helices \eqref{helix_general} increases independent of those signs, we
see that when $\gamma>0$ the $x-y$ components of the helices are traversed in 
a clockwise (CW) manner and the helices are left-handed (LH), while when 
$\gamma <0$ the in-plane motion is CCW, and the helices are right-handed.

We are now in a position to describe quantitatively the helical trajectories
of swimming {\it Chlamydomonas} in the absence of photostimulation.
From the estimated angular
speed $\hat{\omega}_1\sim 1\,$s$^{-1}$ in \eqref{hatomega1}, the typical value 
$\vert\omega_3\vert \sim 10\,$s$^{-1}$ and swimming speed $u\sim 100\,\upmu$m/s, we find 
$R_h\sim 1\,\upmu$m and $P_h\sim 60\,\upmu$m, both
of which agree well with the classic study of swimming trajectories 
(Fig.~6 of \cite{Ruffer1985}), which show the helical radius is a small
fraction of the body diameter and the pitch is $\sim 6$ diameters.

Next we describe in detail the helical trajectories adopted
by cells in steady-state swimming, either toward the 
light during positive phototaxis, or away from it during negative phototaxis.  
For such motions, the relevant angular rotations are $\omega_3$ and the
intrinsic speed $\hat{\omega}_1$.
As noted earlier \cite{Schaller1997}, 
there are four possible
configurations to be considered on the basis of the sense of rotation around ${\bf \hat{e}}_1$ as
determined by {\it cis} dominance ($\hat{\omega}_1 >0$) or {\it trans} dominance ($\hat{\omega}_1<0$).  
In both cases the relationship
between the helix parameters $\gamma$ and $\theta_0$ is%
\begin{subequations}
\begin{align}
    \hat{\omega}_1&=-\gamma \sin\theta_0 \\
    \omega_3&=-\gamma \cos\theta_0,%
\end{align}
\label{omegas}%
\end{subequations}
with $\omega_3<0$ in both cases.
In {\it trans} dominance, $\hat{\omega}_1<0$ and a solution of \eqref{omegas}
has $0\le\theta_0\le \pi/2$, whereas for {\it cis} dominance, $\hat{\omega}_1>0$ and
$-\pi/2\le\theta_0\le 0$.  Thus, 
\begin{equation}
    \theta_0=\tan^{-1}(\hat{\omega}_1/\omega_3), \ \ \ 
    \gamma =\sqrt{\hat{\omega}_1^2+\omega_3^2}.
\end{equation} 
Setting $\tilde{t}=\gamma t$, we obtain the helical trajectories%
    \begin{subequations}
    \begin{align}
      {\bf r}(t)&=\pm R_h\left[\cos\tilde{t}\,{\bf \hat{e}}_x
    -\sin\tilde{t}\,{\bf \hat{e}}_y\right]
    +\frac{P_h \tilde{t}}{2\pi}{\bf \hat{e}}_z, \\
    {\bf \hat{e}}_1&=\cos\theta_0\left[\sin\tilde{t}\,{\bf \hat{e}}_x 
    +\cos\tilde{t}\,{\bf \hat{e}}_y\right]\pm\vert\sin\theta_0\vert\, {\bf\hat {e}}_z, \\
    {\bf \hat{e}}_2&=-\cos\tilde{t}\,{\bf \hat{e}}_x 
    + \sin\tilde{t}\,{\bf \hat{e}}_y.
\end{align}
\label{helix_vectors}%
\end{subequations}
for {\it trans} (+) and {\it cis} ($-$) dominance.

We can now express quantitatively features regarding the eyespot orientation with respect to the 
helical trajectory that have been remarked
on qualitatively \cite{Schaller1997}.  While there is some
variability in the eyespot location, it is 
typically in the equatorial plane defined by $\hat{\bf e}_1$ and $\hat{\bf e}_2$, 
approximately midway between the two.  We take it to lie at an angle 
$\kappa\in [0,\pi/2]$ with respect to
$\hat{\bf e}_2$, such that the outward normal ${\bf \hat{o}}$ to the eyespot is
\begin{equation}
    {\bf \hat{o}}=\sin\kappa\, {\bf \hat{e}}_1 + \cos\kappa\,{\bf \hat{e}}_2.
    \label{eyespotangle}
\end{equation}
The outward normal vectors to the helix cylinder are 
$\hat{\bf n}=\pm\left(\cos\tilde{t}\,\hat{\bf e}_x-\sin\tilde{t}\,\hat{\bf e}_y\right)$, 
so the projection of the eyespot normal on $\hat{\bf n}$ have the time-independent values
\begin{equation}
    \hat{\bf n}\cdot \hat{\bf o}=\begin{cases} -\cos\kappa<0; & {\it trans}\,{\rm dominance}, \\
                          +\cos\kappa>0; & {\it cis}\, {\rm dominance}.
                          \end{cases}
\label{eyespotprojections}
\end{equation}
Thus, for any $\kappa\in[0,\pi/2]$, the eyespot points to the {\it inside} ({\it outside}) of the helix 
for {\it trans} ({\it cis}) dominance.
This confirms the general rule that any given body-fixed spot
on a rigid body executing helical motion due to constant rotations about its axes has 
a time-independent orientation with respect to the helix.  
When $\kappa=0$, $\hat{\bf o}=\hat{\bf e}_2$, which points to the {\it cis} flagellum, and
we see that the dominant flagellum is always 
on the {\it outside} of the helix.

If light shines long some direction $\hat{\bm \ell}$, its
projection on the eyespot is $-\hat{\bm \ell}\cdot \hat{\bf o}$; for
light shining down the helical axis, $\hat{\bm \ell}=-\hat{\bf e}_z$, then
the projections in the two cases are 
\begin{equation}
    -\hat{\bm \ell}\cdot \hat{\bf o}=\begin{cases} +\sin\kappa\,\sin\theta_0 > 0; & {\it trans}\,{\rm dominance}, \\
                          -\sin\kappa\,\vert\sin\theta_0\vert <0; & {\it cis}\, {\rm dominance}.
                          \end{cases}
\label{helixprojections}
\end{equation}

\begin{table}
\caption{Left-handed helical swimming of {\it Chlamydomonas}.}
\begin{ruledtabular}
\begin{tabular}{ccccc}
dominant &  & swimming relative & eyespot & eyespot\\
flagellum & $\hat{\omega}_1$ & to light source & orientation & status\\
\hline 
{\it cis} & $+$ & toward &  outside & shade  \\
{\it cis} & $+$ & away & outside & light \\
{\it trans} & $-$ & toward & inside & light \\
{\it trans} & $-$ & away & inside & shade\\
\end{tabular}
\end{ruledtabular}
\label{table2}
\end{table}

\begin{figure}[t]
\includegraphics[width=0.98\columnwidth]{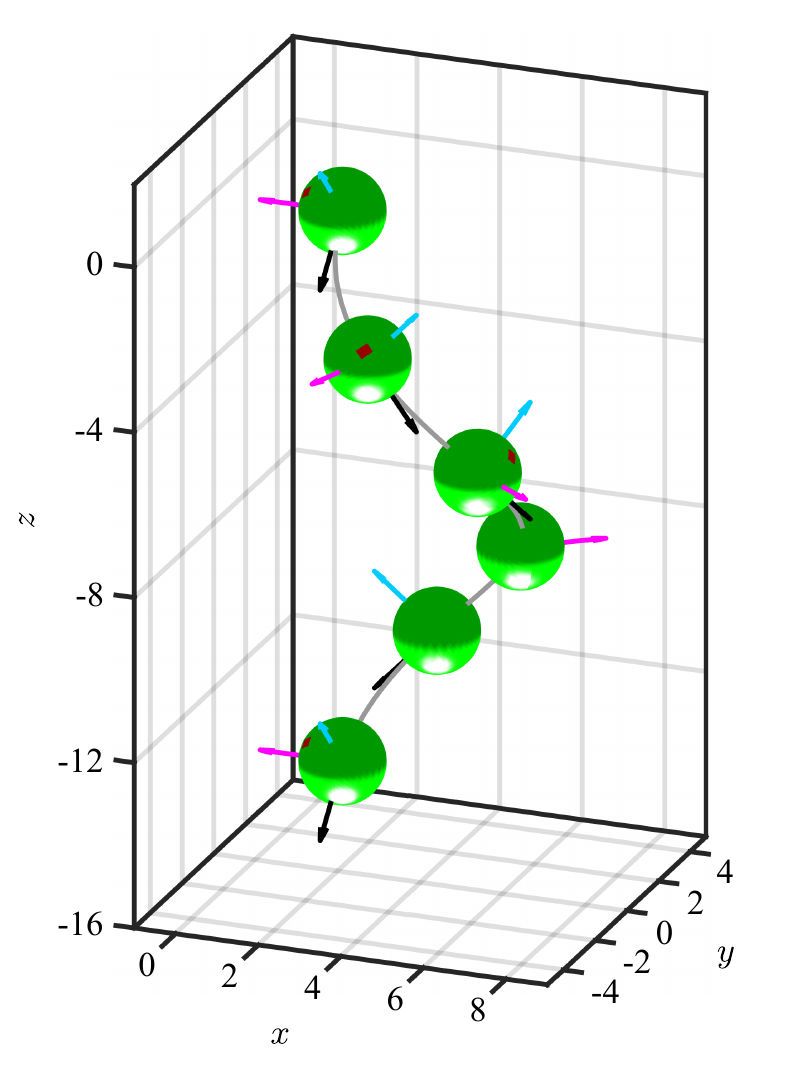}
\caption{Helical trajectory of a \textit{cis}-dominant cell which is swimming in a direction aligned with the light source located at the bottom. Principal body rotation axes are depicted as cyan, magenta and black arrows for $\bf{{\hat{e}}}_1$, $\bf{{\hat{e}}}_2$ and $\bf{{\hat{e}}}_3$ axes respectively. Eyespot is always located in the shading hemisphere of the cell body and is pointing outwards from the helix, as is the \textit{cis} flagellum (not shown).}
\label{fig14}
\end{figure}

These various situations are summarized in Table \ref{table2} and shown in Fig.~\ref{fig14}.
As remarked many years ago, {\it trans} dominance holds in the case of negative phototaxis, and
{\it cis} dominance in positive phototaxis.  The conclusion from Table \ref{table2} is that
when negatively phototactic cells swim away from the light or positively phototactic cells swim
toward the light their eyespots are shaded, and thus the ``stable" state is one minimal signal.
so that
when the first of these conditions holds during negative phototaxis,
whereas the second occurs for positive phototaxis.
Note that in the degenerate case $\theta_0=0$, when the helix reduces to a straight line, the 
projections \eqref{helixprojections} vanish, so a cell moving precisely opposite its desired direction would
receive no signal to turn. A small helical component to the motion eliminates this singular case.

\subsection{Phototactic steering with adaptive dynamics}
\label{photosteering}

Now we merge the adaptive photoresponse dynamics with the kinematics of rigid body motion.  
The dynamics 
for the evolution of the Euler angles in the limit $\omega_2=0$ is obtained from  
\eqref{Eulerangledynamics}, yielding%
\begin{subequations}
\begin{align}
\dot{\phi} &= \omega_1\frac{\sin\psi}{\sin\theta}, \\
\dot{\theta} &= \omega_1\cos\psi, \\
\dot{\psi} &= \omega_3-\omega_1\frac{\sin\psi\cos\theta}{\sin\theta}, 
\label{Euler_intermediate}%
\end{align}
\end{subequations}
Given the assumption $\omega_2=0$, these are exact.  As we take $\omega_3$ to be a constant
associated with a given species of {\it Chlamydomonas}, it remains only
to incorporate the dynamics of $\omega_1$ and the forward swimming speed to have a 
complete description of trajectories.  The angular speed $\omega_1$ is the
sum of intrinsic and phototactic contributions,
\begin{equation}
    \omega_1=\hat{\omega}_1+\omega_1^p,
\end{equation}
where $\omega_1^p$ is described by the adaptive model.

It is natural to adopt rescalings based on the fundamental ``clock" provided by the
spinning of {\it Chlamydomonas} about $\hat{\bf e}_3$.   Recalling that $\omega_3<0$, 
these are 
\begin{align}
    T=&\vert\omega_3\vert t, \ \ \ \ \hat{P}=\frac{\hat{\omega}_1}{\vert\omega_3\vert}, 
    \ \ \ \ P=\frac{\omega_1^p}{\vert\omega_3\vert}, \ \ \ \ \alpha=\vert\omega_3\vert\tau_a,\nonumber \\
    \beta&=\vert\omega_3\vert \tau_r, \ \ \ \ H=\frac{h}{\vert\omega_3\vert}, 
    \ \ \ \ S=\frac{s}{\vert\omega_3\vert}.
    \end{align}

To incorporate the photoresponse, 
the light signal at the eyespot must be expressed in terms of the Euler angles.  Henceforth we
specialize to the case in which the a light source shines in the $x-y$ plane
along the negative $x$-axis, so $\hat{\bm \ell}=-\hat{\bf e}_x$, and 
the normalized projected light intensity $J=-\hat{\bm \ell}\cdot \hat{\bf o}$ on the 
eyespot can be written as
\begin{equation}
   J=\sin\left(\kappa-\psi\right)\cos\phi
   -\cos\left(\kappa-\psi\right)\cos\theta\sin\phi.
   \label{Jdef}
\end{equation}
We assume for simplicity that eyespot shading is perfect, so that
the signal sensed by the eyespot is
\begin{equation}
    S= P^* J {\cal H}(J),
    \label{signaldef}
\end{equation}
where $P^*=\omega_1^*/\vert\omega_3\vert$ is negative (positive) for  
positive (negative) phototaxis, and 
${\cal H}$ is the Heaviside function.

\begin{figure*}[t]
\includegraphics[width=1.90\columnwidth]{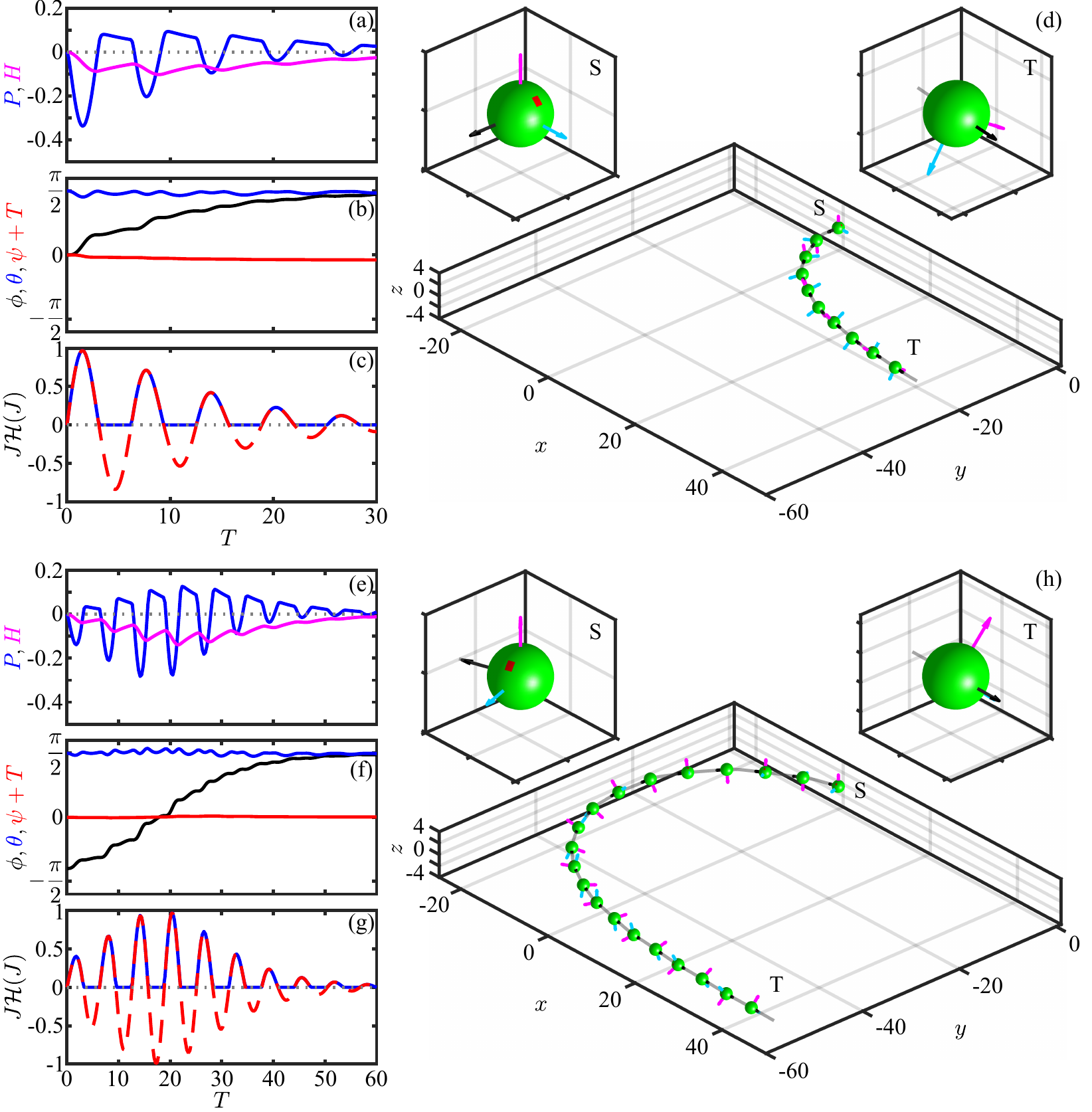}
\caption{Positive phototaxis from model for non-helical swimming. 
(a-c) Time evolution of adaptive variables, Euler angles, and
eyespot signal during a phototurn, i.e.~from swimming orthogonal to the light to moving directly
toward it. (d) Trajectory of the turn showing initial (S) and final (T) orientations of the cell (also in magnified insets). (e-h) Analogous to panels above but with dynamics starting from an 
orientation facing nearly away from the light. Parameters are $\hat{P}=0$, $P^*=-0.4$, $\alpha=7$, 
$\beta=0.14$, and $U=2$, with the eyespot along $\hat{\bf e}_2$ (i.e.~$\kappa=0$ but shown in canonical position), and $\tau_d=0$. Color scheme of cell's principal axes is same as in Fig.~\ref{fig14}.}
\label{fig15}
\end{figure*}

\begin{figure*}[t]
\includegraphics[width=1.95\columnwidth]{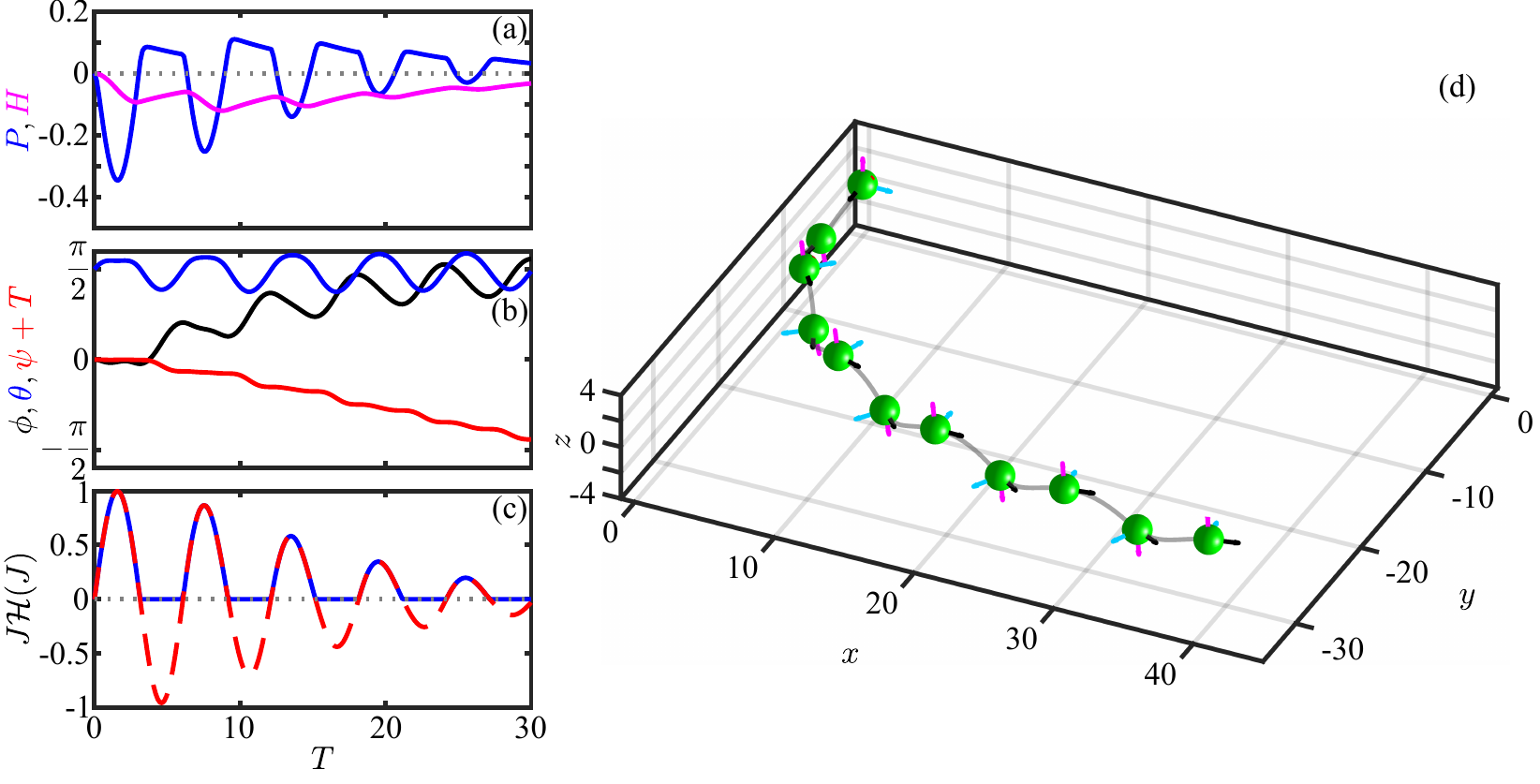}
\caption{Positive phototaxis with helical swimming. 
As in Fig.~\ref{fig15} but with $\hat{P}=0.3$.}
\label{fig16}
\end{figure*}

With these rescalings, the dynamics reduces to 
\begin{subequations}
\begin{align}
\phi_T &= \left(\hat{P}+P\right)\frac{\sin\psi}{\sin\theta}, \label{fd1} \\
\theta_T &= \left(\hat{P}+P\right)\cos\psi,\label{fd2} \\
\psi_T &= -1- \left(\hat{P}+P\right)\frac{\sin\psi\cos\theta}{\sin\theta},\label{fd3} \\
\beta P_T &= S-H-P, \label{fd4}\\
\alpha H_T &= S-H.\label{fd5}
\end{align}
\label{fulldynamicsscaled}%
\end{subequations}
These five ODEs, supplemented with the signal definition in \eqref{Jdef} and 
\eqref{signaldef},
constitute a closed system.
To obtain the swimming trajectory, we use the cell body radius $R$ to 
define the scaled position vector
${\bf R}={\bf r}/R$, so that the dynamics
${\bf r}_t =u\hat{\bf e}_3$ becomes
\begin{equation}
    {\bf R}_T=U\left\{\sin\theta\left[\sin\phi\,{\bf \hat{e}}_x 
    -\cos\phi\,{\bf \hat{e}}_y\right]
    +\cos\theta\, {\bf\hat {e}}_z\right\},
    \label{tangent_general}
\end{equation}
where $U=u/(R\vert\omega_3\vert)$ is the scaled swimming speed.  For typical parameter
values ($u=100\,$\textmu m/s, $R\sim 5\,$\textmu m, and $f_r=1.6\,$Hz), we find $U\sim 2$.
Given $(\theta(T),\phi(T))$, we integrate \eqref{tangent_general} forward from some
origin ${\bf R}(0)$ to obtain ${\bf R}(T)$, and use the triplet $(\theta(T),\phi(T),\psi(T))$ and
the matrix ${\bf A}$, the inverse of $\tilde{\bf A}$ in \eqref{tildeA}, to obtain 
$\hat{\bf e}_1(T)$ and $\hat{\bf e}_2(T)$.

An important structural feature of the dynamics 
is its partitioning into sub-dynamics for the Euler angles 
and the flagellar
response.  The connection between the two is provided by the response variable $P(T)$,
via the signal $S$, such that any
other model for the response (for example, one incorporating distinct dynamics for 
the {\it cis} and {\it trans} flagella), or the signal (including only partial 
eyespot directionality, or cell body lensing) can be substituted for the adaptive  
dynamics with perfect shading.  

The model \eqref{fulldynamicsscaled} has 
$4$ main parameters: $\hat{P}$ determines the unstimulated swimming helix, 
$P^*$ sets the maximum photoresponse turn rate, and $\alpha$ and $\beta$ describe
the adaptive dynamics.   Additional parameters are the eyespot angle 
$\kappa$ \eqref{eyespotangle} and  time delay $\tau_d$.
To gain insight, we first adopt the simplification that the eyespot 
vector $\hat{\bf o}$ is along $\hat{\bf e}_2$ ($\kappa=0$), set $\tau_d=0$,
and solve the initial value problem in which a cell starts swimming in the
$x-y$ plane ($\theta(0)=\pi/2$) along the 
direction $-\hat{\bf e}_y$ ($\phi(0)=\pi/2$) with its eyespot orthogonal to the light
($\psi(0)=0$; $\hat{\bf e}_1=\hat{\bf e}_x$ and $\hat{\bf e}_2=\hat{\bf e}_z$) 
and about to rotate into the light.  Figure \ref{fig15} shows the results of 
numerical solution of the model for the non-helical case $\hat{P}=0$, with
$P^*=-0.4$ for positive phototaxis, $\alpha=7$, $\beta=0.15$ and $U=2$.  We see 
in Fig.~\ref{fig15}(a) how the initially large photoresponse when the cell is
orthogonal to the light decreases with each subsequent half turn as the angle $\phi$
evolves toward $\pi/2$ [Fig.~\ref{fig15}(b)].  The signal at the eyespot [Fig.~\ref{fig15}(c)],
is a half-wave rectified sinusoid with an exponentially decreasing amplitude.
Note that for this non-helical case the Euler angle $\theta$ remains very close to
$\pi/2$ during the entire phototurn, indicating that the swimmer remains nearly in the
$x-y$ plane throughout the trajectory.  

If the initial condition is nearly opposite to the light direction, the model shows that
the cell can execute a complete phototurn (Fig.~\ref{fig15}(e).
In the language of Schaller, et al. \cite{Schaller1997},
since the eyespot is ``raked backward" when the cell swims toward the light, and
the eyespot is shaded, when the cell swims away from the light it picks up a signal
and can execute a full $\pi$ turn to reach the light.

Next we include helicity in the base trajectory, setting $\hat{P}=0.3$.  
In the absence of phototactic stimulation this value leads to helical motion with
a ratio of helix radius to pitch of $R_h/P_h=\hat{P}/2\pi\simeq 1/20$, a value
considerably larger than to that seen experimentally \cite{Ruffer1985}, but useful for the purposes of
illustration.
The phototurn dynamics shown in Fig.~\ref{fig16} exhibits the same qualitative 
features seen without helicity, albeit with much more pronounced
oscillations in the evolution of the Euler angles, particularly of $\phi$ and $\theta$.
Averaged over the helical path the overall trajectory is similar to that without helicity,
and does not deviate significantly from the $x-y$ plane.

To make analytical progress in quantifying a phototurn, we use the simplifications
that are seen in Fig.~\ref{fig15} for the non-helical case.  First, we neglect
the small deviations of $\theta$ from $\pi/2$, and simply set $\theta=\pi/2$.  Second, we note
that the time evolution of $\psi$ is dominated by rotations around 
$\hat{\bf e}_3$, and thus we assume $\psi=-T$.  This yields a simplified model in which the remaining
Euler angle $\phi$ is driven by the cell spinning, subject to the adaptive dynamics.%
\begin{subequations}
\begin{align}
\phi_T &= -P\sin T,\label{fulldynamics1_a} \\
\beta P_T &= S-H-P, \\
\alpha H_T &= S-H.
\end{align}
\label{fulldynamics1}%
\end{subequations}
where we allow for a general eyespot location, using 
$J=\cos\phi\sin(T+\kappa)$, and thus
\begin{equation}
    S=P^*\cos\phi\sin \left(T+\kappa\right)\, 
    {\cal H}\left(\sin \left(T+\kappa\right)\right).
    \label{eq:simple}
\end{equation}

As it takes a number of half-periods of body rotation to execute a turn, we can consider the angle 
$\phi$ to be approximately constant during each half turn $n$ ($n=1,2,\ldots$) at the value we label $\phi_n$.
For any fixed $\phi_n$, the signal $S$ is simply a HWRS of amplitude
$P^*\cos\phi_n$.
We explore two approaches to finding the evolution of $\phi_n$: (i) a {\it quasi-equilibrium}
one in which the steady-state response of the adaptive system to an
oscillating signal is used to estimate $P$, and (ii) a {\it non-equilibrium} one
in which the response is the solution to an initial-value problem.

In the first approximation, we decompose the HWRS eyespot signal
\eqref{eq:simple} 
into a Fourier series,
\begin{align}
    S=P^*\cos\phi_n\,\Bigl[&\frac{1}{\pi}
    +\frac{1}{2}\sin\left(T+\kappa\right)\nonumber \\
    &-\frac{2}{\pi}\sum_{n=1}^{\infty}\frac{\cos(2n\left(T+\kappa\right))}{(2n)^2-1}
    \Bigr].
\end{align}
From the linearity of the adaptive model it follows that 
each term in this series produces an independent response with magnitude $G$ and phase shift $\chi$ appropriate
to its frequency $n\omega_3$, for $n=0,1,2,\ldots$.  Since the magnitude $G$ in \eqref{gainphase_a}
vanishes at zero frequency, the contributing terms in the photoresponse are
\begin{align}
    P=P^*&\cos\phi_n\Biggl\{\frac{1}{2}G(\omega_3)\sin\left[T+\kappa+\chi\left(\omega_3\right)\right] \\
    &-\frac{2}{\pi}\sum_{n=1}^{\infty}\frac{G(2n\omega_3)}{(2n)^2-1}\cos\left[2n\left(T+\kappa\right)
    +\chi\left(2n\omega_3\right)\right]
    \Biggr\}\nonumber
\end{align}
The first term dominates, as it is at the same frequency as the r.h.s. of the equation of motion $\phi_T=-P\sin T$.  
Keeping only this term, we integrate 
\begin{equation}
   \phi_T\simeq -\frac{1}{2} P^*G(\omega_3)\cos\phi_n \sin T\sin(T+\kappa+\chi)
   \label{phi_eqn}
\end{equation}
over one half period ($T=\pi$) and obtain the iterated map
\begin{equation}
\phi_{n+1} = \phi_n+\xi\cos{\phi_n},
\label{itermap}
\end{equation}
where 
\begin{equation}
    \xi=-\frac{\pi}{4}P^* G(\omega_3)\cos\left[\kappa+\chi(\omega_3)\right],
    \label{xidefine}
\end{equation}
with $\xi>0$ for $P^*<0$ in positive phototaxis and $\xi<0$ for negative phototaxis.  
An alternative approach involves the direct integration of the equations of motion over
each half-turn.  The lengthy algebra for this is given in Appendix \ref{app:map}, where
one finds a map analogous to \eqref{itermap}, but with an $n$-dependent factor $\xi_n$
that converges for large $n$ to that in \eqref{xidefine}.  Supplementary Video 2 \cite{SM} illustrates the cell reorientation dynamics under this map.

\begin{figure}[t]
\includegraphics[width=0.98\columnwidth]{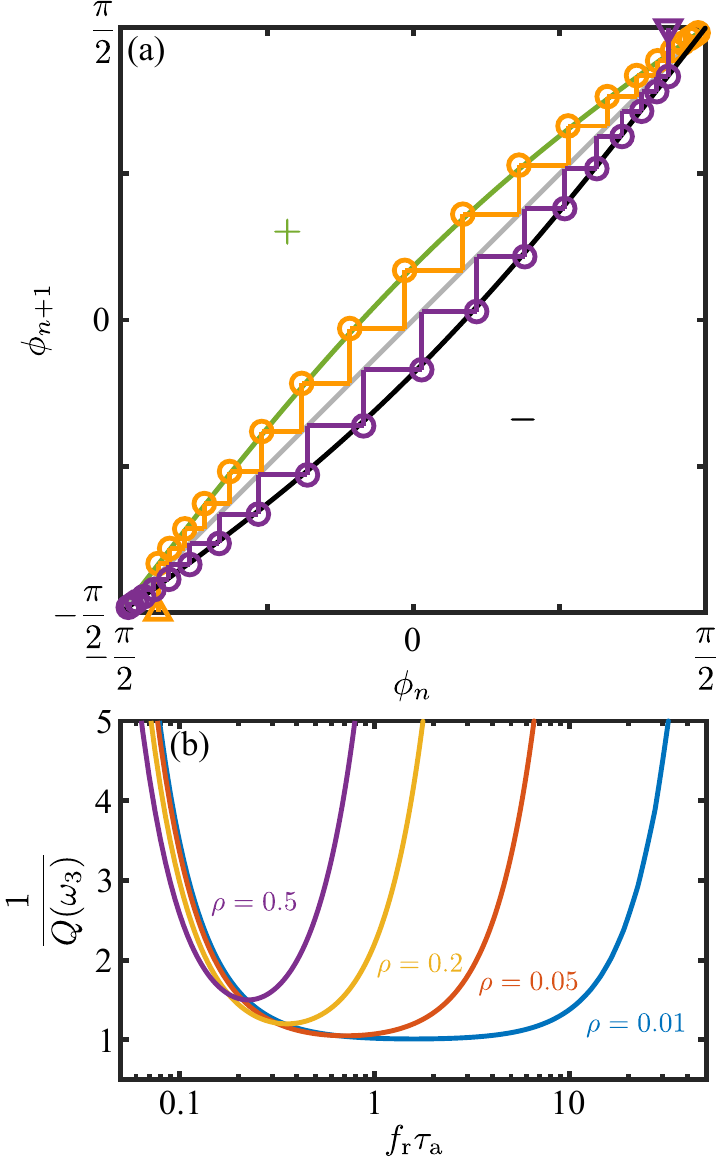}
\caption{Iterated map of the reorientation model.  (a) Cobwebbing of iterations starting from
$\phi$ near $-\pi/2$ for positive phototaxis (upper branch in yellow) and near $+\pi/2$ for negative phototaxis 
(lower branch in purple), as indicated by triangles, for light shining toward $-\hat{\bf e}_x$. Values of $\xi=\mp0.29$ are used which are calculated based on \eqref{xidefine} using values from Table \ref{table3}. (b) Response factor $1/Q$ in \eqref{turnnumber} for
number of half-turns needed for alignment
as a function of tuning parameter, for various values of $\rho$ as indicated.}
\label{fig17}
\end{figure}

The iterated map \eqref{itermap} has fixed points at $\phi_\pm=\pm \pi/2$.
Linearizing about those values by setting $\phi_n=\pm \pi/2+\delta\phi_n$, we obtain
$\delta\phi_{n+1}=(1\mp\xi)\delta\phi_n$ and thus $\delta\phi_n \propto (1\mp \xi)^n \delta\phi_1$.
Hence: (i) the angle $+\pi/2$ is
stable for positive phototaxis when $0\le \xi \le 2$ and becomes unstable for $\xi>2$, while it is
unstable for negative phototaxis ($\xi<0$). (ii) the angle $-\pi/2$ is unstable for positive phototaxis
for any $\xi>0$, while it is stable for negative phototaxis in the range $-2\le \xi\le 0$ and
unstable for $\xi < -2$.  Thus, positively phototactic cells orient toward $+\pi/2$ and negatively
phototactic cells orient toward $-\pi/2$, except for values of $\vert\xi\vert>2$.  These 
exceptional cases correspond to
peak angular speeds $\omega_1^* > 4\vert\omega_3\vert/3\sim 13\,$s$^{-1}$.

Figure \ref{fig17}(a) shows the iterated map \eqref{itermap} for both positive and negative
phototaxis.  In the usual manner of interpreting such maps, the ``cobwebbing" of successive iterations
shows clearly how the orientation $+\pi/2$ is the global attractor for positive phototaxis, and 
$\phi=-\pi/2$ is that for negative phototaxis. 
When $\vert\xi\vert$ is small, the approach to the stable fixed point is exponential,
$\delta\phi \sim \exp(-n/N)$, where $N=N_0/Q(\omega_3)$, with
\begin{equation}
    Q(\omega_3)=G(\omega_3)\cos\left(\kappa+\chi_0-\delta\right),  \ \ \ \ N_0=\frac{4}{\pi \vert P^*\vert},
    \label{turnnumber}
\end{equation}
is the characteristic number half-turns needed for alignment.  The number $N_0$ reflects the 
bare scaling with the maximum rotation rate around $\hat{\bf e}_1$. 
For the typical value $P^*\simeq 0.1$ we have $N_0\sim 6$.
The presence of the gain $G$ in denominator
in \eqref{turnnumber} embodies the effect of tuning between the adaptation timescale and the 
rotation rate around $\hat{\bf e}_3$, while the term $\cos\left(\kappa+\chi_0-\gamma\right)$ 
captures the feature discussed in frequency response studies in Sec. \ref{sec:adaptive}, namely that 
the flagellar asymmetries have maximum effect (and thus $Q$ is maximized) when the negative 
phase shift $\chi_0-\delta$ offsets the eyespot location). 
Figure \ref{fig17}(b) shows the dependence of $N/N_0$ on the scaled relaxation time $f_r\tau_a$ for
various values of $\rho$.  For the experimentally observed range $\rho\simeq 0.02-0.1$ there is a
wide minimum of $N/N_0$ around $f_r\tau_a\sim 1$.  This relationship confirms
the role of tuning in the dynamics of phototaxis, but also shows the
robustness of the processes involved.

Returning to the evolution equation \eqref{phi_eqn} for $\phi$, we can also average the term 
$\sin T\sin(T+\kappa+\chi)$ over one complete cycle to obtain the approximate evolution equation
\begin{equation}
    \phi_T=\frac{1}{4}\vert P^* \vert Q(\omega_3)\cos\phi. 
    \label{phi_eqn_final}
\end{equation}
For positive phototaxis and with $\phi(0)=0$, the solution to this ODE can be expressed in 
unrescaled units as 
\begin{equation}
	\Phi(t) = 2\tan^{-1}\left(e^{t/\tau_x}\right),
	\label{reorient}
\end{equation}
where $\Phi=\pi/2+\phi$, with $\Phi(0)=\pi/2, \Phi(\infty)=\pi$, and the characteristic time in physical units is
\begin{align}
    \tau_x&=\frac{4}{\vert \omega_1^* \vert Q(\omega_3)}\nonumber \\
    &=\frac{4\tilde{\zeta}_r/\vert{\mathcal T}_{\mathrm{p}}^*\vert}{G(\omega_3)\cos\left(\kappa+\chi_0-\delta\right)} .
    \label{reorient_tau}
\end{align}
This is a central result of our analysis, in that it relates the time scale for reorientation during
a phototurn to the magnitude and dynamics of the
transient flagellar asymmetries during the photoresponse.  As discussed above, the
function $Q(\omega_3)$ embodies the optimality of the response\textemdash in terms of
the tuning between the rotational frequency and the adaptation time, and 
the phase delay and eyespot position\textemdash but also captures the robustness of the
response through the broad minimum in $Q$ as a function of both frequency
and eyespot position.

\begin{figure}[t]
\includegraphics[width=0.98\columnwidth]{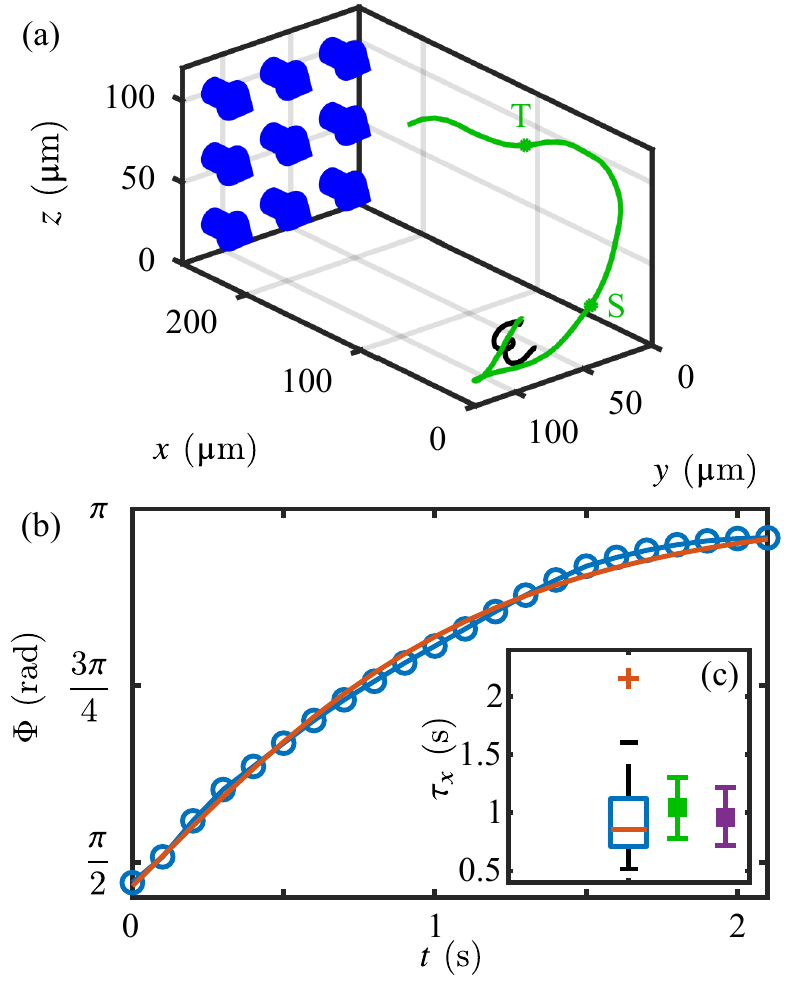}
\caption{Phototactic swimmers tracked in three-dimensions.
(a) A U-turn: trajectory in black is prior to light stimulation, that in green is afterwords. 
Blue arrows indicate direction of light. 
The cropped trajectory used for fitting the reorientation dynamics is bounded by the points S and T.
(b) Dynamics of the reorientation angle $\Phi$ (blue) for the cropped trajectory fitted using \eqref{reorient}.
(c) Box plot of distribution of fitted $\tau_x$ with estimates, steady-state (green) and nonequilibrium (purple), derived from micropipette experiments.}
\label{fig18}
\end{figure}

Using the 3D tracking system described in Sec. \ref{sec:methods}, we analyzed 
$6$ pairs of movies, within which we tracked $283$ trajectories with duration greater than $10\,$s and 
which included the trigger frame.  From those, $44$ showed both positive phototaxis and included a full turn 
to $\phi = \pi/2$ as shown in Fig.~\ref{fig18}(a) and Supplementary Video 3 \cite{SM}.
These trajectories were cropped to include any points for which $-\pi/2 \leqslant \phi 
\leqslant \pi/2$ and which could then be fitted to \eqref{reorient} to determine the experimental
time constant $\tau_x$.
The boxplot in Fig.~\ref{fig18} shows the experimental mean value $\tau_x=0.96 \pm 0.36\,$s (also in Table \ref{table3}).  
This can be compared to the estimates obtained within the steady-state approximation \eqref{reorient_tau} and the 
transient analysis (Appendices \ref{app:map} and \ref{app:tau_x}), both of which are based on the mean and standard error value of the peak flagellar 
phototorque of $-35.0\pm 8.8\,$pN$\cdot\upmu$m, the mean values of the adaptive flagellar response time-scales 
($\tau_r = 0.018\,$s and $\tau_a = 0.764\,$s in Table \ref{table3}) and the effective drag coefficient $\tilde{\zeta}_r$.
The steady-state estimate of $\tau_x$ is $1.04\pm 0.26\,$s (with $Q/4 = 0.243$), while 
the transient estimate is $0.96\pm 0.24\,$s (with the nonequilibrium counterpart $\bar{\mathcal{Q}} = 0.265$ 
in \eqref{reorient_tau_transient}). This agreement provides strong validation of the model of adaptive phototaxis
developed here.  

\begin{table}[t]
\caption{Time scales that define the reorientation dynamics. Experimental data are from the
present study except for the cell body rotation frequency.}
\begin{ruledtabular}
\begin{tabular}{cccc}
\textbf{Quantity} & \textbf{Symbol} & \textbf{Mean} ${\bm{\pm}}$ \textbf{SD}\\
\hline 
adaptation time  & $\tau_{\mathrm{a}}$ & $0.764\,\pm\,0.190\,$s\\
response time  & $\tau_{\mathrm{r}}$ & $0.018\,\pm\,0.009\,$s\\
time delay & $\tau_{\mathrm{d}}$ & $0.028\,\pm\,0.011\,$s \\  
cell body rotation frequency \cite{Choudhary}& $f_{\mathrm{r}}$ & $1.67\,\pm\,0.35\,$Hz\\
reorientation time scale & $\tau_{\mathrm{x}}$ & $0.96\,\pm\,0.36\,$s\\
\end{tabular}
\end{ruledtabular}
\label{table3}
\end{table}

\section{Discussion}

This study has achieved three goals: the development of methods to capture flagellar 
photoresponses at high spatio-temporal resolution, the estimate of torques generated during 
these responses and the measuremnt of relevant 
biochemical time scales that underlie 
phototaxis, and the integration of this information into a mathematical model to describe 
accurately the phototactic turning of \textit{Chlamydomonas}.
In developing a theory for phototurns, our work also puts on a more systematic mathematical 
foundation qualitative 
arguments \cite{Schaller1997} for the stability of phototactic trajectories 
based on eyespot orientation in both positive and negative phototaxis. 

We have emphasized that rather than seek to develop a maximally detailed
model of the dynamics of individual flagellar responses involved in phototaxis, we aimed to provide,
in the context of one simple microscopic model,
a multiscale analysis of the connection between such responses and the phototactic trajectories in a
manner than can be easily generalized. Thus we obtain from experiment the values for microscopic and 
macroscopic time scales, as shown in Table \ref{table3}, and derive relation between them, culminating 
in \eqref{reorient_tau} (and \eqref{reorient_tau_transient}). 

This analysis highlights the dual issues of optimatility and
robustness.  As noted in the introduction, the former was 
first addressed using a paralyzed-flagella mutant strain (\textit{pf14}) and an electrophysiological approach 
on a bulk sample by \cite{Yoshimura2001}. In those experiments, a suspension of immotile cells was exposed to 
an oscillating light stimulus (wavelength $500\,$nm) and the resulting photoreceptor current was measured in a 
cuvette attached to two platinum electrodes. The experiment using relatively high light intensities observed 
a frequency response peak of $1.6\,$Hz when stimulated with $\approx$160\,\textmu E$\udot$s$^{-1}\udot$m$^{-2}$ 
and a frequency response peak of $3.1\,$Hz when stimulated with $\approx$ 40 \textmu E$\udot$s$^{-1}\udot$m$^{-2}$.
The former observation is in very good agreement with our results in Fig.~\ref{fig13} 
(peak response at $\simeq 2\,$Hz), even though we used 
light stimulus intensities of $\approx$1 \textmu E$\udot$s$^{-1}\udot$m$^{-2}$. We have not seen any evidence of 
cells having flagellar photoresponse dynamics that would corroborate the latter result of $3.1\,$Hz and this 
is a matter open to further study.

In addition, this study has addressed issues relating to past observations.  
With respect to the lag time $\tau_d$ of the photoresponse, we have measured by detailed
study of the flagellar waveforms a value of 
$28\,\pm\,11\,$ms that is very similar to the value 
$30-40\,$ms observed earlier \cite{Ruffer1991}. In addition, we have shown through 
the adaptive dynamics that the peak flagellar response is at a larger total delay
time $t^*$ given by \eqref{eq:t_max} that corresponds accurately to the time
between the eyespot receiving a light signal and the alignment of the flagellar
beat plane with the light.  Analysis of the phototactic model reveals that such
tuning shortens the time for phototactic alignment.

Regarding the amount of light necessary for a flagellar photoresponse appropriate to positive 
phototaxis, we have converged, 
through trial and error, to $\approx$1 \textmu E$\udot$s$^{-1}\udot$m$^{-2}$ at a wavelength of $470\,$nm. 
While this value is much lower than in other photoresponse experiments \citep{Josef2005} where 
$\approx$60 \textmu E$\udot$s$^{-1}\udot$m$^{-2}$ were used at a longer wavelength ($543\,$nm), 
it is consistent with the sensitivity profile of channelrhodopsin-2 \citep{Sineshchekov_two_2002}. 
More detailed studies 
of the wavelength sensitivity of the flagellar photoresponse should be carried out in order to reveal any 
possible wavelength dependencies of quantities such as the time constants $\tau_{\mathrm{r}}$ and $\tau_{\mathrm{a}}$.
Our work has addressed the relationship between the stimulus and the photoresponse 
of \textit{Chlamydomonas} using an adaptive model that has perhaps the minimum number of parameters appropriate
to the problem, each corresponding to a physical process. Attempts to derive similar relationships 
between stimulus and photoresponse \citep{Josef2006} used linear system analysis. The result of such a 
signal-processing oriented method usually includes a much larger number of parameters necessary 
for the description of the system, without necessarily corresponding to any obvious measurable physical quantities.

The evolutionary perspective that we emphasized in the
introduction, culminating in the results presented in 
Fig.~\ref{fig2}, points to several areas for future
work.  Chief among them is an 
understanding of the biochemical origin of the
response and adaptive timescales of the photoresponse,
in light of genomic information available on the
various species.
Flagellar and phototaxis mutants will likely be important
in unravelling whether these time scales 
are associated with the
axoneme directly or arise from coupling to 
cytoplasmic components.  Additionally, we anticipate
that directed evolution experiments such as
those already applied to {\it Chlamydomonas} 
\cite{Ratcliff}
can yield important information on the dynamics of
phototaxis.  For example, is it possible to evolve cells that exhibit faster phototaxis, and if so,
which aspect of the light response changes? For the multicellular green algae,
these kinds of experiments may also impact on the 
organization of somatic cells within the 
extracellular matrix, which has been shown to
exhibit significant variability \cite{Yunker}.

Another aspect for future investigation sits within the general are of control theory;
the adaptive phototaxis mechanism that is common to the
Volvocine algae, and to other systems such as 
{\it Euglena} \cite{Tsang}, is one in which a chemomechanical 
system achieves a fixed point by evolving in time
so as to null out a periodic signal. Two natural questions
arise from this observation. First, what evolutionary
pathways may have led to this behavior? 
Second, are there lessons for control theory in general and perhaps even for autonomous vehicles 
in particular
that can be deduced from this navigational strategy?

We close by emphasizing that the flagellar photoresponse -- and by extension phototaxis -- is a 
complex biological process encompassing many variables, and that in addition to the short-term responses to light stimulation studied here there
are issues of long-term adaptation to darkness or phototactic light that have only recently have begun to be addressed \citep{Arrieta}.
Together with the dynamics of phototaxis in concentrated suspensions \cite{Brunet}, these are
important issues for further work.

\begin{acknowledgments}

We thank Pierre A. Haas and Eric Lauga for very useful discussions and a critical 
reading of the manuscript, Kirsty Y. 
Wan for sharing code from previous work on flagellar tracking, David-Page Croft, Colin Hitch, 
John Milton, and Paul Mitton or technical support, and Ali Ghareeb for assistance with the 
fiber coupling apparatus.  This work was supported in part by Wellcome Grant 207510/Z/17/Z.
For the purpose of open access, the authors have 
applied a CC BY public copyright license to any Author Accepted Manuscript version 
arising from this submission.  Additional support
was provided by
Grant No. 7523 from the Marine Microbiology Initiative of the Gordon and Betty
Moore Foundation, and EPSRC Established Career Fellowship EP/M017982/1.
\end{acknowledgments}

\appendix
\begin{widetext}


\section{Details of the iterated map from solution of the initial value problem} 
\label{app:map}

Here we provide details of the derivation of the iterated map for phototurns based
on explicit solution of the initial value problem for the adaptive response.  For conciseness we fix
the eyespot position at $\kappa=0$ and set the time delay $\tau_d=0$.  We start from the 
dynamics \eqref{fulldynamics1}, rewritten for each full turn $n\ge 0$ as
\begin{equation}
	\alpha H^{(n)}_T + H^{(n)} = \begin{cases}
						P^*\cos\phi \sin{T}; & n\pi \leqslant T < (n+1)\pi,\, n\,{\rm even} \\
						0; & n\pi \leqslant T < (n+1)\pi,\, n\,{\rm odd}
					 \end{cases}
\end{equation}
and
\begin{equation}
    \beta P^{(n)}_T + P^{(n)} = \begin{cases}
						P^*\cos\phi\sin{T} - H^{(n)}; & n\pi \leqslant T < (n+1)\pi,\, n\,{\rm even} \\
						-H^{(n)}; & n\pi \leqslant T < (n+1)\pi,\, n\,{\rm odd}.
					 \end{cases}
\end{equation}
Solving this in a piecewise fashion we obtain
\begin{equation}
H^{(n)} = P^*\cos\phi \frac{\alpha}{1+\alpha^2}\times 
\begin{cases} C_n e^{-(T-n\pi)/\alpha} 
+ \frac{1}{\alpha}\sin{T} - \cos{T} ; & n\pi \leqslant T < (n+1)\pi,\, n\,{\rm even} \\
	C_n e^{-(T-n\pi)/\alpha}; & n\pi \leqslant T < (n+1)\pi, n\,{\rm odd}.
\end{cases}
\end{equation}
where $C_n=(1-r^{n+1})/(1-r)$ and $r=e^{-\pi/\alpha}$.  Continuity of $H$ at the end of each light 
interval can be verified by noting that $H^{(n)}(T=(n+1)\pi)\propto rC_n+1$ for even $n$, 
while $H^{(n)}(T=n\pi)\propto C_{n+1}$ for the subsequent odd $n$, and observing that $1+rC_n=C_{n+1}$.

The solution for the photoresponse variable can be expressed as $P^{(n)}=P^*\cos\phi\, \tilde{P}^{(n)}$, where
\begin{equation}
\tilde{P}^{(n)} =  \begin{cases}
	\Lambda_1 D_n e^{-(T-n\pi)/\beta} - \Lambda_2C_n e^{-(T-n\pi)/\alpha}
	+ \Lambda_3\sin{T} + \Lambda_4\cos{T} ; & n\pi \leqslant T < (n+1)\pi, n\,{\rm even} \\
	\Lambda_1 D_n e^{-(T-n\pi)/\beta} - \Lambda_2C_n e^{-(T-n\pi)/\alpha}; & n\pi \leqslant T < (n+1)\pi, n\,{\rm odd} \label{eq:Ptilde_def}
\end{cases}
\end{equation}
with  $D_n=(1-q^{n+1})/(1-q)$, $q=e^{-\pi/\beta}$, and
\begin{equation}
\Lambda_1= \frac{\alpha\beta}{(1+\beta^2)(\alpha-\beta)}, \ \ 
\Lambda_2= \frac{\alpha^2}{(1+\alpha^2)(\alpha-\beta)}, \ \
\Lambda_3= \frac{\alpha(\alpha+\beta)}{(1+\beta^2)(1+\alpha^2)}, \ \ 
\Lambda_4= \frac{\alpha(1-\alpha\beta)}{(1+\beta^2)(1+\alpha^2)}. 
\end{equation}

\end{widetext}

\begin{figure}[H]
	\includegraphics[width=0.85\columnwidth]{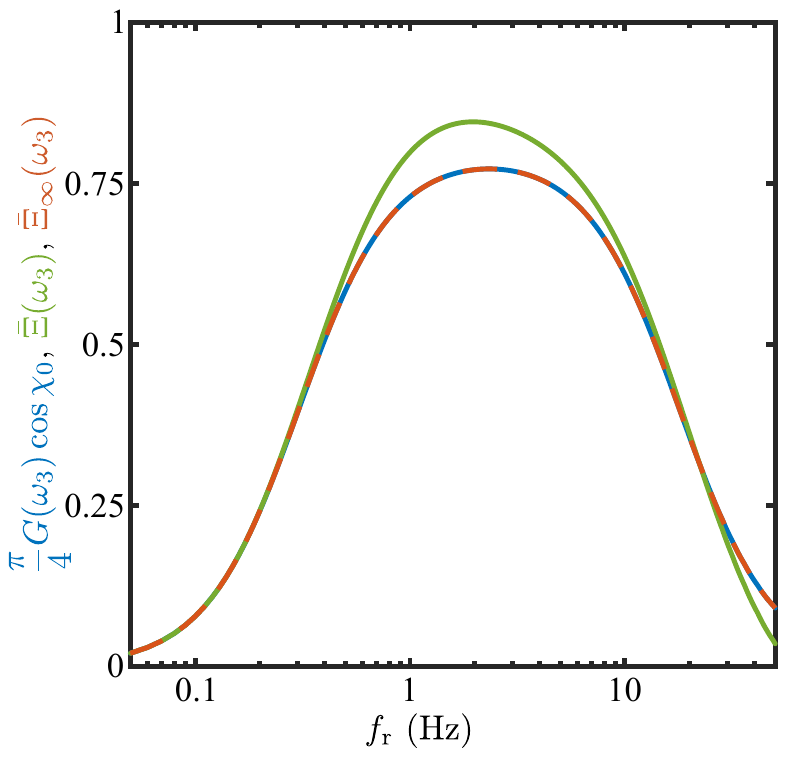}
	\caption{Response functions. 
 For  $\kappa=\tau_d=0$, the graph compares the steady-state response 
 function $(\pi/4)G(\omega_3)\cos{\chi_0}$ in \eqref{gainphase} (dashed blue), the initial average 
 response $\bar{\Xi}(\omega_3)$ in \eqref{barxi} (green) and the $n\to \infty$ limit 
 of the transient response $\bar{\Xi}_{\infty}(\omega_3)$ (dashed red), as functions of $f_r=\omega_3/2\pi$.
	}
\label{fig19}
\end{figure}

Since $n$ represents the number of half-turns, 
with even(odd) values for the illuminated(shaded)
periods, we integrate for each value of $n \geqslant 0$ 
to obtain
\begin{equation}
\phi_{n+1} = \phi_{n}-P^*\cos\phi\int_{n\pi}^{(n+1)\pi} \tilde{P}^{(n)} \sin{T} dT.
\end{equation}
This has the form of \eqref{itermap}, but with an
$n$-dependent $\xi_n$,%
\begin{equation}
\xi_n = -P^*\Xi_n(\omega_3)\label{eq:xi_n}
\end{equation}
where
\begin{equation}
	\Xi_n(\omega_3) = \begin{cases}
	   \displaystyle \beta^2\Lambda_1 D_n(q+1) \\
       \displaystyle \qquad - \alpha^2\Lambda_2 C_n(r+1) +\frac{\pi}{2}\Lambda_3, &n\,{\rm even} \\
       \displaystyle -\beta^2\Lambda_1 D_n(q+1) + \alpha^2\Lambda_2 C_n(r+1), &n\,{\rm odd}.
\end{cases}
\label{eq:Xi_n}
\end{equation}
From the general structure of the iterated map, it is clear that the larger is $\xi_n$ the larger the
angular change within a given half-turn.  It is of interest then to consider the average 
$\bar{\Xi} = \left(\Xi_0 + \Xi_1\right)/2$ over the first two half turns, which
gives the average coefficient
\begin{equation}
\bar{\xi} = -P^*\bar{\Xi}(\omega_3).
\label{barxi}
\end{equation}
The quantity $\bar{\Xi}$ can be interpreted as the initial photoresponse function analogous to the
steady-state response embodied in the amplitude $G(\omega_3)$ and phase $\chi_0$ in \eqref{gainphase}.
The functions $G(\omega_3)\cos\chi_0(\omega_3)$ and $\bar{\Xi}$ are compared in Fig.~\ref{fig19}, where we 
see that the transient 
response function $\bar{\Xi}$ is about $10\%$ higher at its peak,
a feature that can be attributed
to the fact that the hidden variable $H$ has not yet built up to its steady value.  But the two
functions are otherwise remarkably similar, indicating the accuracy of the steady-state approximation.

More generally, the coefficients $\Xi_n$ exhibit an oscillating decay with $n$, converging as $n\to \infty$ to
\begin{align}
	\Xi_{\infty}(\omega_3) = \begin{cases}
	\displaystyle \beta^2\Lambda_1\frac{1+q}{1-q} - \alpha^2\Lambda_2 \frac{1+r}{1-r} +
\frac{\pi}{2}\Lambda_3, &n\,{\rm even}\nonumber \\
\displaystyle -\beta^2\Lambda_1\frac{1+q}{1-q} + \alpha^2\Lambda_2 \frac{1+r}{1-r}, &n\,{\rm odd}.
\end{cases}
\end{align}
The connection to the steady-state approximation is obtained by 
considering the average  
$\bar{\Xi}_{\infty}(\omega_3)$
over the light and dark cycles (hence over even and odd values of $n$), where one finds  $\bar{\Xi}_{\infty}(\omega_3) = (\pi/4)\Lambda_3$, or
\begin{equation}
	\bar{\Xi}_{\infty}(\omega_3) = \frac{\pi}{4}G(\omega_3)\cos(\chi_0),
\end{equation}
completely consistent with the steady-state analysis \eqref{xidefine}.

In the animation shown in Supplementary Video 2 \cite{SM} 
of the cell reorientation dynamics we evolve the
iterated map using the $\left\{\xi_n\right\}$ and 
linearly interpolate between the values $\phi_n$ to obtain
a smooth function of time.

\begin{figure}[H]
	\includegraphics[width=0.85\columnwidth]{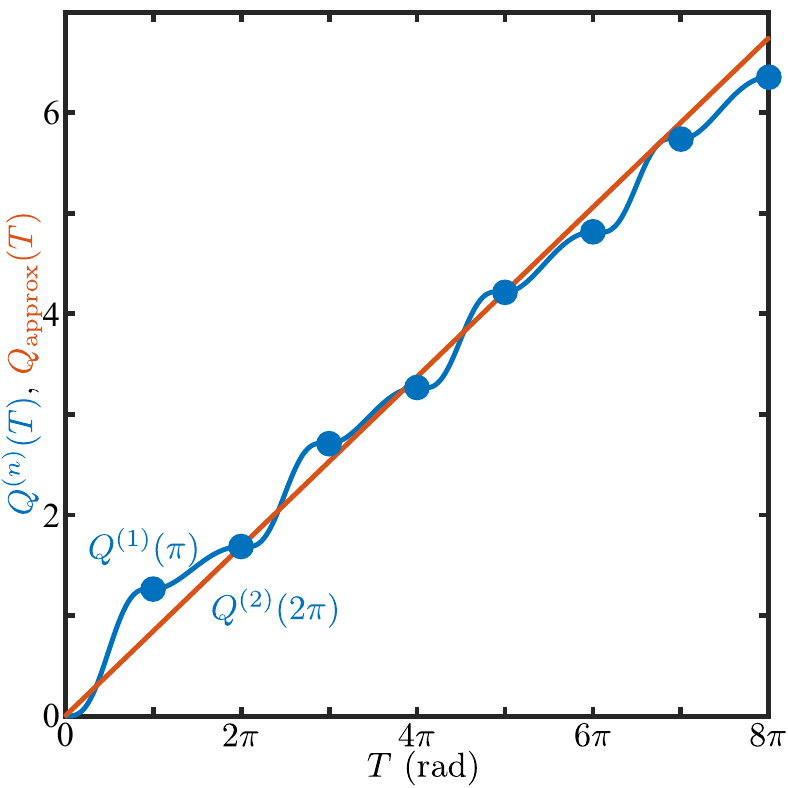}
	\caption{The functions $Q^{(n)}(T)$. Approximating $Q^{(n)}(T)$ 
	as a straight line $Q_{\mathrm{approx}}(T)=\bar{\mathcal{Q}}T$. 
	Both $Q^{(n)}(t)$ and $\bar{\mathcal{Q}}$ were computed with 
	the experimentally derived $\tau_r = 0.009$ s and 
	$\tau_a = 0.520$ s. The value of $f_r$  was taken to be 1.67 Hz. 
	The value of $\bar{\mathcal{Q}}(\omega_3, \tau_r, \tau_a)$ 
	was calculated to be 0.265.
	}
\label{fig20}
\end{figure}


\section{Details of continuous model}
\label{app:tau_x}

\begin{figure*}[t]
\includegraphics[width=1.7\columnwidth]{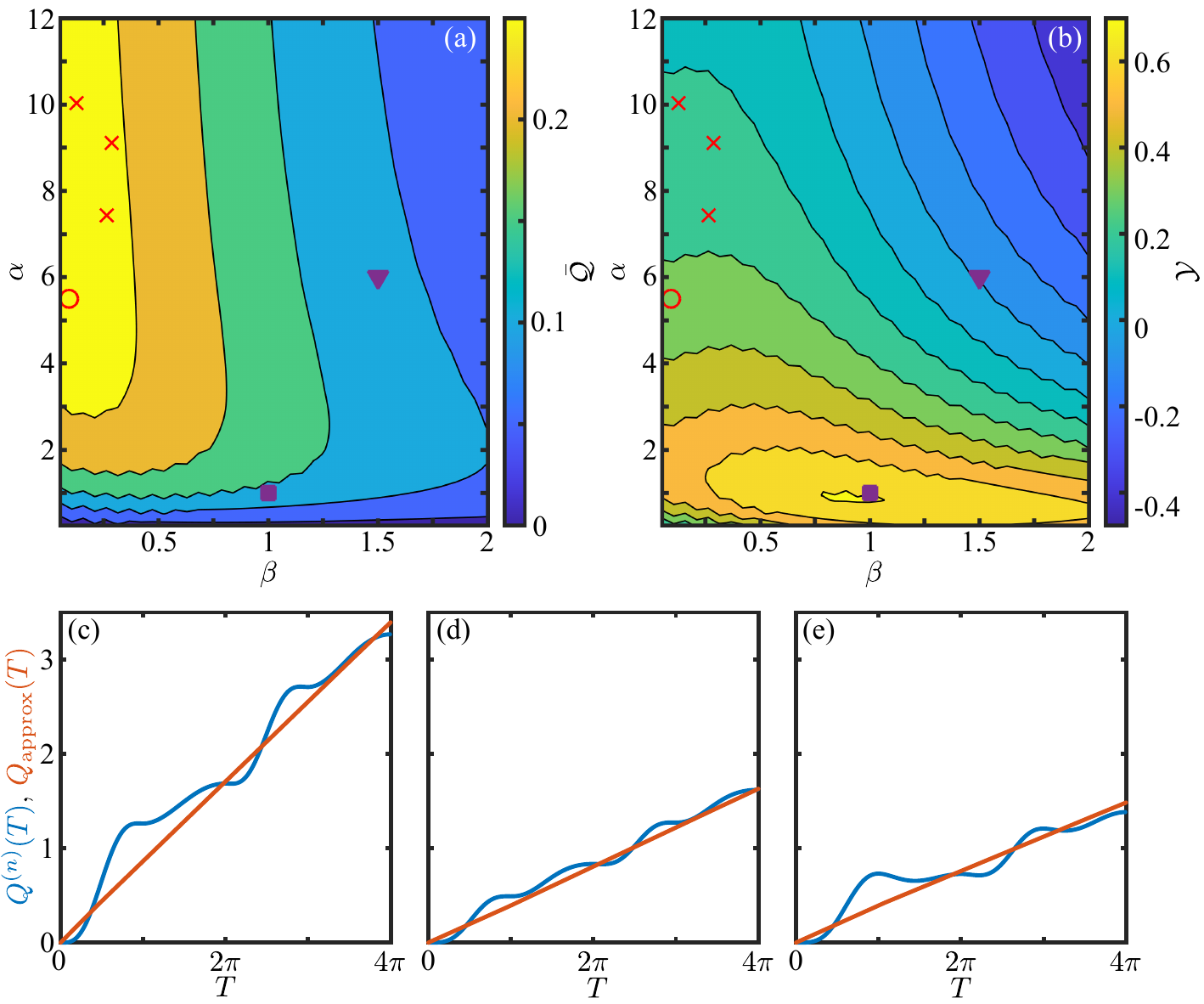}
\caption{Contour plot maps of $\bar{\mathcal{Q}}$ (a) and $\mathcal{Y}$ (b) with ($\alpha$,$\beta$) pairs acquired from micropipette experiments shown with red markers (step-up as ``x'' and frequency as ``o"). (c-e) Plots of $Q^{(n)}(T)$ (blue line) and linear approximation $Q_{\mathrm{approx}}(T)=\bar{\mathcal{Q}}T$ (red line) for three ($\alpha$,$\beta$) value pairs shown in (a) and (b) as circle, square and triangle respectively. (c) is based on experimental data (open red circle),  (d) corresponds to $\mathcal{Y} \approx 0.7$ (solid purple square) and (e) corresponds to $\mathcal{Y} \approx 0$ (solid purple triangle).}
\label{fig21}
\end{figure*}
\begin{widetext}
Using the solution of the initial value problem we can compute the continuous approximation
to the evolution equation for $\phi$ by integrating over
the fast photoresponse variables within a turn.
From the governing equation
$\phi_T = -P^*\cos\phi\, \tilde{P}^{(n)}(T)\sin{T}$, we
obtain 
\begin{equation}
	\Phi(T) = 2\tan^{-1}\left(e^{-P^* Q^{(n)}(T)}\right), 
\end{equation} 
where
\begin{equation}
	Q^{(n)}(T) = \int_0^T \tilde{P}^{(n)}(T') \sin{T'}{\rm d}T'.\label{eq:phi_from_Q}
\end{equation}

From \eqref{eq:Ptilde_def} we find
\begin{equation}
Q^{(n)}(T) = \begin{cases}
\displaystyle -\beta\Lambda_1 D_n e^{-\frac{T-n\pi}{\beta}}(\sin{T}+\beta\cos{T}) + \alpha\Lambda_2 C_n e^{-\frac{T-n\pi}{\alpha}}(\sin{T}+\alpha\cos{T}) \\
\displaystyle \qquad + \Lambda_3\left(\frac{T}{2} - \frac{\sin{2T}}{4}\right) + \Lambda_4\frac{\sin^2{T}}{2} - \Lambda_3\frac{n\pi}{4} + \beta^2 \Lambda_1 - \alpha^2 \Lambda_2 & n\pi \leqslant T < (n+1)\pi, n\,{\rm even}\\
	\displaystyle -\beta\Lambda_1 D_n e^{-\frac{T-n\pi}{\beta}}(\sin{T}+\beta\cos{T}) + \alpha\Lambda_2 C_n e^{-\frac{T-n\pi}{\alpha}}(\sin{T}+\alpha\cos{T}) \\
	\displaystyle \qquad + \Lambda_3\frac{(n+1)\pi}{4} & n\pi \leqslant T < (n+1)\pi, n\,{\rm odd}.
\end{cases}
\end{equation}
As shown in Figure \ref{fig20}, the function $Q^{(n)}(T)$ typically increases monotonically with 
$T$, exhibiting small oscillations around an interpolant that grows nearly linearly with time.
These magnitudes of these oscillations vary between the light and dark halves of each turn. To
quantify this asymmetry we compute
the values $Q^{(n)}(n\pi)$ at the start of each half turn,
\begin{equation}
Q^{(n)}(n\pi) = \begin{cases}
\beta^2\Lambda_1 (1-D_n) - \alpha^2\Lambda_2 (1 - C_n) + \Lambda_3\frac{n\pi}{4}, & 
\text {\qquad$n$\,even }  \nonumber\\
	\displaystyle \beta^2\Lambda_1 D_n  - \alpha^2\Lambda_2 C_n + \Lambda_3\frac{(n+1)\pi}{4} & 
\text {\qquad$n$\,odd }
\end{cases}
\end{equation}
\end{widetext}
and the gradients $\displaystyle \mathcal{Q}_n = [Q^{(n+1)}((n+1)\pi) - Q^{(n)}(n\pi)]/\pi$ of line segments
connecting the node. 
One can easily show that $\mathcal{Q}_n = \Xi_n/\pi$.
The light-dark variation of these slopes serves as a measure of the smoothness
of the reorientation dynamics, and from the first two values $\mathcal{Q}_0$ and 
$\mathcal{Q}_1$ we define two relevant quantities: the {\it strength} of the initial
response as measured by the average of the slopes of the first two line segments 
$\bar{\mathcal{Q}} = \left(\mathcal{Q}_0 + \mathcal{Q}_1\right)/2 = \bar{\Xi}/\pi$, 
and its {\it smoothness}, as measured by the ratio $\mathcal{Y} = \mathcal{Q}_1/\mathcal{Q}_0$. 

If, as in Fig.~\ref{fig20}, we approximate $Q^{(n)}(T)$ by the line $\bar{\mathcal{Q}}T$, then
the reorientation dynamics \eqref{eq:phi_from_Q} takes the simple form
\begin{equation}
	\Phi(T) = 2\tan^{-1}\left(e^{-P^* \bar{\mathcal{Q}} T}\right),
	\label{eq:phi_from_xi}
\end{equation}
from which we identify the characteristic relaxation time
$\tau_x$ (in physical units) analogous to \eqref{reorient_tau},
\begin{equation}
    \tau_x=\frac{1}{\vert \omega_1^*\vert \bar{\mathcal{Q}}} = \frac{\tilde{\zeta_r}}{\vert \mathcal{T}_{\mathrm{p}}^*\vert \bar{\mathcal{Q}}}.
    \label{reorient_tau_transient}
\end{equation}

Finally, we explore the space of reorientation dynamics by probing $Q^{(n)}(T)$ through its dependency 
on parameters $\alpha$ and $\beta$. Our strategy is to observe how the quantities 
$\bar{\mathcal{Q}}(\alpha,\beta)$ (Fig.~\ref{fig21}(a)) and $\mathcal{Y}(\alpha,\beta)$ 
(Fig.~\ref{fig21}(b)), which are also functions of $\alpha$ and $\beta$, and essentially describe the 
curve's shape, vary. Firstly, we make the observation that the ($\alpha$,$\beta$) 
pairs acquired from micropipette experiments (step-up and frequency response; Fig.~\ref{fig10}(b)) lie 
in the high-slope area ($\bar{\mathcal{Q}} = 0.27$) of the $\bar{\mathcal{Q}}(\alpha,\beta)$ 
function (Fig.~\ref{fig21}(a) and Fig.~\ref{fig21}(c)). We also observe that the same data lie 
in an area of relatively moderate symmetry ($\mathcal{Y} \approx 0.34$) of the 
$\mathcal{Y}(\alpha,\beta)$ function (red markers in Fig.~\ref{fig21}(b)), as opposed to 
the extreme cases of highest symmetry i.e.~$\mathcal{Y} \approx 0.7$ (solid purple square in 
Fig.~\ref{fig21}(b) and Fig.~\ref{fig21}(d)) and lowest symmetry i.e.~$\mathcal{Y} \approx 0$ 
(solid purple triangle in Fig.~\ref{fig21}(b) and Fig.~\ref{fig21}(e)).

\end{document}